\documentclass[12pt]{article}
\usepackage{latexsym,epsfig,graphicx,amsmath,amssymb,amscd,undertilde,multirow,chicago,psfrag,paralist,dsfont}
\usepackage[titletoc]{appendix}
\newcommand{\thetavec}{{\boldsymbol{\theta}}}

\newcommand{\wvec}{{\boldsymbol{w}}}

\newcommand{\E}{\mathbf{E}}
\newcommand{\Es}{\mathbf{E}_{\ast}}

\newcommand{\thetavechat}{\widehat{\thetavec}}

\newcommand{\wh}{\widehat}

\newtheorem{result}{Result}
\newcommand{\Xvec}{\boldsymbol{X}}

\newcommand{\xvec}{\boldsymbol{x}}

\newcommand{\lbar}{\bar{l}}
\newcommand{\lbars}{\bar{l}^{\ast}{}}
\newcommand{\thetavechats}{\wh\thetavec^{\ast}}
\newcommand{\V}{\textrm{Var}}
\newcommand{\Vs}{\textrm{Var}_{\ast}}
\newcommand{\N}{\textrm{N}}

\begin{document}

\title{Applications of the Fractional-Random-Weight Bootstrap}

\author{Chris Gotwalt\\
JMP Division, SAS\\\
Research Triangle, NC 12345\\
\and
Li Xu and Yili Hong\\
Department of Statistics\\
Virginia Tech\\
Blacksburg, VA 24061\\
\and
William Q. Meeker\\
Department of Statistics\\
Iowa State University\\
Ames, IA 50011\\
}

\date{\today}

\maketitle
\begin{abstract}
The bootstrap, based on resampling, has, for several decades, been a widely used method for computing confidence intervals for applications where no exact method is available and when sample
sizes are not large enough to be able to rely on easy-to-compute large-sample approximate methods,
such a Wald (normal-approximation) confidence intervals. Simulation based bootstrap intervals have been proven useful in that their actual coverage probabilities are close to the nominal confidence level in small samples. Small samples analytical approximations such as the Wald method, however, tend to have coverage probabilities that greatly exceed the nominal confidence level. There are, however, many applications where the resampling bootstrap method cannot be used. These include situations where the data are heavily
censored, logistic regression when the success response is a rare event or where there is insufficient
mixing of successes and failures across the explanatory variable(s), and designed experiments where the
number of parameters is close to the number of observations. The thing that these three situations have
in common is that there may be a substantial proportion of the resamples where is not possible to
estimate all of the parameters in the model. This paper reviews the fractional-random-weight
bootstrap method and demonstrates how it can be used to avoid these problems and construct
confidence intervals. For the examples, it is seen that the fractional-random-weight
bootstrap method is easy to use and has advantages over the resampling method in many challenging applications.

\textbf{Key Words:} Censored data, Confidence interval, Design of experiments, Prediction interval, Resampling,
Variable selection.
\end{abstract}

\newpage

\section{Introduction}\label{secINTRO}
\subsection{Bootstrap Background}
The bootstrap is a popular statistical tool that is used, primarily, to obtain improved inferences such as
approximate confidence intervals and approximate prediction intervals that have coverage probabilities
that are close to the nominal confidence level. Bootstrapping is a set of procedures for sampling from the distribution of an estimator that most commonly employs various data generation and augmentation procedures to create new data sets from which new individual samples of the estimator are computed. These samples of the estimators can then be used for many purposes, including approximate confidence and prediction intervals that have more desirable inferential properties than their more commonly used deterministic counterparts. With modern computing technology (hardware and
software) bootstrap methods are easy to implement and use. Bootstrap procedures can be applied even
in situations where classical theory offers little or no guidance about how to compute trustworthy
confidence interval. Generally, there are only minimal regularity conditions (such as a finite variance and
a certain degree of smoothness) that are needed to make bootstrap methods work well. Technical details of bootstrap methods can be found in classical references such as \citeN{h92}, \citeN{EfronTibshirani1993}, \citeN{st95}, and \citeN{dh97}.

There are many different types of bootstrap procedures which can be broadly partitioned into
nonparametric and parametric. Nonparametric bootstrap procedures require no assumptions about the
shape of the underlying data-generating probability distribution. Most bootstrap procedures do assume
sampling from a continuous or approximately continuous distribution.
Bootstrapping is most commonly done via a Monte Carlo simulation. The most common approach is to
generate a sequence of new data sets using resampling. In resampling, each new data set is generated
by sampling the rows of the original data with replacement. This approach for generating the bootstrap
samples is nonparametric because no assumption about the shape of the underlying distribution is
required.

Bootstrap samples can also be generated by assuming a particular parametric distribution and
simulating from that distribution. In applications where censoring or truncation is involved, censoring
and truncation must be done in a manner that mimics what was done in the original data-generating
process. For example, if censoring is random, then a model for the censoring variable needs to be used
in the parametric simulation. Often details about how data were censored are not known or is
complicated and in such situations, the nonparametric resampling method is much easier to implement.
After each bootstrap data set is generated, the statistical procedure (e.g., model fitting and computation
of point estimates and in some cases standard errors) is applied to the bootstrap dataset and results are
stored. This bootstrap-sample generation/estimation procedure is repeated a number of times (e.g.,
2,500 times) and then the saved results are processed to make inferences (e.g., construct confidence
intervals). There are many different ways to use bootstrap samples to compute a confidence interval
(e.g., simple percentile, bias-corrected percentile, bias-corrected and accelerated, percentile-$t$ intervals).

\subsection{The Idea of Data Weights}
In many data analysis applications, it is convenient to put weights (also known as frequencies or counts) on
observations. For example, binary data such as 0010001000100010001 are usually replaced with counts
of the number of zeros and ones. Weights are frequently used in life test data. For example, the data
typically consist of the failure times (which all have weight 1 unless there are ties---often caused by
failures being recorded at discrete inspection times) plus the number of units that survived a 1,000-hour
test. Weights are also used in data compression where data are binned and the weights indicate the
number of observations in each bin (e.g., as displayed in a histogram). Also, when observations have known
non-constant variances, it is appropriate to use weights that are inversely proportional to the variance
of each observation. The resampling bootstrap method of generating bootstrap samples can also be
viewed as data with random \textit{integer} weights. That is, each observation has a weight indicating the
number of times it was drawn in the resample. We will give an explicit example of this in the next
section.

Many statistical estimation methods allow the specification of weights or frequencies. For example,
consider a data set $\xvec_n= (x_1, x_2,\ldots, x_n)$ with corresponding weights $\wvec_n=(w_1 , w_2,\ldots, w_n)$. Then estimates of the mean ($\mu$)
and variance ($\sigma^2$) can be computed from
$$
\widehat{\mu}=\frac{1}{\sum_{i=1}^{n}w_i}\sum_{i=1}^nw_ix_i,\quad\text{	and	}\quad \widehat{\sigma}^{2}=\frac{1}{\sum_{i=1}^{n}w_i}\sum_{i=1}^nw_i(x_i-\widehat{\mu})^2.
$$
There are similar equations for more general weighted least squares for linear regression models. More
generally, suppose we have a dataset $\xvec_n=(x_1$, $x_2$, $\dots$, $x_n$) with corresponding weights $\wvec_n=(w_1, w_2,\dots$, $w_n)$,
where each $x_i$ may contain information such as a response, explanatory variables, and censoring or
truncation indicators for observation $i$. Then the weighted likelihood is
$$
L(\thetavec;\xvec_n, \wvec_n)=\mathcal{C}\prod_{i=1}^n \left[L_i(\thetavec;x_i)\right]^{w_i},
$$
where $\thetavec$ is a general notation for the unknown parameters, $\mathcal{C}$ is a constant that does not related to $\thetavec$, and $L_i(\thetavec;x_i)$ is the likelihood contribution from observation $i$.

In general, we can see that the data weight idea is common in statistical methods and it provides an easy way for computational implementations. The data weight idea also provides an alternative way to understand bootstrap methods. The objective of this paper is to review the random \textit{fractional-weight} bootstrap method and demonstrate how to apply it to applications in which the resampling (integer weight) bootstrap methods tend to not work well. These applications include heavily censored data, logistic regression when the success response is a rare event or where there is insufficient
mixing of successes and failures across the explanatory variable(s), and designed experiments where the number of parameters is close to the number of observations.

\subsection{Literature Review}
Much has been written about the bootstrap methods since their introduction in the late 1970s. For
example, the textbooks by \citeN{EfronTibshirani1993}, and \citeN{dh97} describe bootstrap
theory and methods. The books by \citeN{h92} and \citeN{st95} focus on the theory behind
bootstrap methods. Another notable reference, aimed at teaching bootstrap methods, is \citeN{h15}.

As there are only a handful of articles devoted to it, in spite of its usefulness, the FRW bootstrap sampling method appears to be quite under-appreciated at this point.  We believe the FRW bootstrap method could serve a much larger role in the toolkit of the applied statistician. \citeN{bb95} provide a highly technical presentation of the asymptotic theory of various random-weight methods for generating
bootstrap estimates. They show how to choose the distribution of the random weights by using
Edgeworth expansions. \citeN{cb05} present a generalized bootstrap for which the
traditional resampling and various weighted likelihood and other weighted estimating equation
methods are special cases. \shortciteN{cjw05} apply the FRW bootstrap methods to a recurrent events
application with informative censoring in a semi-parametric model. \citeN{hmm09} apply FRW
bootstrap methods to a prediction interval application involving complicated censoring and truncation.
\citeN{xhm15} use the FRW bootstrap in a prediction application to assess the risk of
future failures.

\subsection{Overview}
The remainder of this paper is organized as follows. Section~\ref{secIFWB} introduces the concept of integer and fractional-weight bootstrap methods.
Section~\ref{secTR} gives some theoretical properties of the FRW bootstrap method. Section~\ref{secACI} provides applications of the FRW bootstrap in confidence intervals, which involves heavily censored field-failure data, current-status censored data, and data that are sensitive in estimating the shape parameter of the generalized gamma distribution. Section~\ref{secAPI} illustrates the application of the FRW bootstrap to compute prediction intervals using an application to predict failure of power transformers. Section~\ref{secADE} describes an example where the FRW bootstrap is used to find an appropriate model to describe the results of a designed experiment. Section~\ref{secCRAFR} provides some concluding remarks and areas for further research. Technical details are given in the appendix.

\section{Integer and Fractional-Weight Bootstrap}\label{secIFWB}
\subsection{Integer-weight Bootstrap}
Under the idea of data weights, the commonly-used resampling bootstrap procedure is equivalent to choosing the weights from a multinomial distribution with uniform probability $1/n$ for each of the
original observations in the sample, where $n$ is the number of observations. That is, the weights $(w_1, \ldots, w_n)'$ follows a uniform multinomial distribution that has a mean 1 and a variance $(n-1)/n$.

As an illustration, the first column of Table~\ref{tab:int.frw.wts} gives data tree volume for 15 loblolly pine trees in units of cubic meters. The
data are a subsample of data analyzed in Chapter 13 of \citeN{mhe17}. The other
three columns give the results of resampling with replacement from the sample of size 15, indicating the
number of times that each tree was selected for each of the three resamples. As described in Section \ref{secINTRO},
in an actual application of the bootstrap the resampling would be done $B$ times, usually on the order of
thousands. Then a weighted estimation method could be applied to each bootstrap resample to obtain
the $B$ bootstrap estimates.

\begin{table}
\begin{center}
\caption{Three integer-weight (on the left) and fractional-weight (on the right) bootstrap samples.}\label{tab:int.frw.wts}
\vspace{1.5ex}
\begin{tabular}{cccccccc}\hline\hline
\multirow{4}{*}{Tree Volume}  & \multicolumn{3}{c}{Uniform} && \multicolumn{3}{c}{Uniform}\\
 & \multicolumn{3}{c}{Multinomial Distribution} && \multicolumn{3}{c}{Dirichlet Distribution}\\
 & \multicolumn{3}{c}{Integer Weights}&& \multicolumn{3}{c}{Continuous Weights}\\\cline{2-4} \cline{6-8}
 & $j=1$ & $j=2$ & $j=3$&&  $j=1$ & $j=2$ & $j=3$\\\hline
 0.149 & 1& 1 &1 && 0.203&	0.485&	1.451   \\
 0.086 & 2& 0 &0 && 0.065&	1.328&	2.062   \\
 0.149 & 3& 0 &0 && 0.629&	1.737&	0.676   \\
 0.194 & 0& 0 &1 && 0.505&	0.953&	0.590   \\
 0.044 & 1& 1 &0 && 0.735&	1.510&	0.580   \\
 0.104 & 1& 1 &1 && 2.543&	0.320&	2.512   \\
 0.156 & 0& 2 &1 && 2.650&	0.714&	1.320   \\
 0.122 & 1& 0 &1 && 0.690&	2.072&	0.650   \\
 0.117 & 0& 3 &2 && 1.095&	0.017&	0.901   \\
 0.079 & 3& 0 &2 && 2.075&	1.344&	0.792   \\
 0.179 & 0& 0 &1 && 0.020&	2.368&	0.061   \\
 0.307 & 0& 7 &0 && 1.947&	0.116&	1.917   \\
 0.049 & 0& 0 &1 && 1.433&	0.633&	0.982   \\
 0.165 & 1& 0 &2 && 0.131&	1.137&	0.212   \\
 0.043 & 2& 0 &2 && 0.279&	0.265&	0.294   \\\hline
 Sum &15& 15&15 &&  15&      15&     15\\\hline\hline
\end{tabular}
\end{center}
\end{table}

%
\subsection{Fractional-weight Bootstrap Samples}\label{sec:fwboot}
Extending the idea of integer weights, the FRW is introduced with continuous weights. In this case, the weight vector $(w_1, \ldots, w_n)'$ is generated from a from a uniform Dirichlet distribution, multiplied by $n$. The probability density function (pdf) of the Dirichlet distribution of order $n$ with parameters $\alpha_1, \ldots, \alpha_n$ is given by
\begin{align}\label{eqn:dirichelet}
f(w_1,\ldots, w_{n}; \alpha_1,\ldots, \alpha_n )=\frac{1}{\mathrm{B}(\alpha_1,\ldots, \alpha_n)} \prod_{i=1}^n w_i^{\alpha_i - 1}, \quad\quad \sum_{i=1}^{n} w_i=1, \quad\quad w_i \ge 0,
\end{align}
and $\mathrm{B}(\alpha_1,\ldots, \alpha_n)$ is the normalizing factor. The uniform Dirichlet distribution is a special case when $\alpha_i=1, i=1,\ldots, n$. The continuous weights, like the integer multinomial resampling weights, will sum to $n$, have expectation 1, but a variance $(n-1)/(n + 1)$. As an illustration, the last three columns of Table~\ref{tab:int.frw.wts} shows the random fractional weights drawn from a uniform Dirichlet distribution, multiplied by $n$.

The FRW bootstrap was first suggested by \citeN{r81}. He called it the Bayesian bootstrap because,
as shown in the paper, estimates computed from the FRW bootstrap samples are draws from a posterior
distribution under a particular relatively diffuse prior distribution. \citeN{nr94} generalized
Rubin's ideas and introduced the weighted likelihood bootstrap, which is easy to implement. \citeN{nr94} also show that the weighted likelihood bootstrap is first order accurate.

Even though these random-weight
bootstrap methods were developed within a nonparametric Bayesian framework, they also apply to
non-Bayesian and parametric inference problems, as will be illustrated in the examples in this paper.
There are statistically valid alternative methods to generate the random fractional weights. In particular,
\citeN{jyw01} show that FRW bootstrap estimators have good properties if positive independent and
identically distributed (iid) weights are generated from a continuous distribution that has a mean and
standard deviation equal to one (e.g., an exponential distribution with mean one). Then the sum of the weights is a random variable with expectation $n$.

The FRW or Bayesian bootstrap is also known as: the random-weight bootstrap, the weighted likelihood bootstrap, the weighted bootstrap, and the perturbation bootstrap. Operationally, the FRW bootstrap samples are used in the same way as the resampling bootstrap
samples. Like resampling, the method is nonparametric. There are, however, important advantages of
using the FRW bootstrap in certain common applications. The advantages arise because all of the
original observations remain in all of the bootstrap samples. In situations where dropping certain
observations from a data set will cause estimation problems, the resampling bootstrap approach will often give poor results or fail altogether. Generally, when using the FRW bootstrap, because all of the original observations remain in the sample, estimation difficulties do not arise.

\section{Theoretical Results}\label{secTR}
In this section, we present some theoretical results, which provide the basis for the statistical inference for the FRW bootstrap. For likelihood based inference, it can be shown that the fractional weights generated from the uniform Dirichlet distribution are equivalent to generating standardized random weights from an exponential distribution with mean one. Let $Z_i, i=1,\ldots,n$ be independent and identically distributed exponential distribution with mean one. Then the random vector
\begin{align}\label{equ:sumton}
\left(\frac{Z_1}{\sum_{i=1}^nZ_i},\ldots,\frac{Z_i}{\sum_{i=1}^nZ_i} \ldots,\frac{Z_n}{\sum_{i=1}^nZ_i}\right)'
\end{align}
has a uniform Dirichlet distribution.

For convenience of the development of the theoretical properties, we use the weights from the exponential distribution with mean one.
Let $X_1,X_2,\ldots, X_n$ be $n$ random iid observations and $\Xvec_n$ a general notation for the collection of the $n$ random observations. The loglikelihood function can be written as
$$\lbar(\thetavec)=\frac{1}{n}\sum_{i=1}^nl_i(\thetavec; X_i),$$
where $l_i(\thetavec; X_i)$ is the contribution for observation $i$ and $\thetavec$ is a general notation for the vector of unknown parameters. The maximum likelihood (ML) estimate $\thetavechat$ is the solution to the first derivative $\lbar'(\thetavec)=\partial \lbar(\thetavec)/\partial\thetavec=0.$
The random weighted loglikelihood is
$$\lbars(\thetavec)=\frac{1}{n}\sum_{i=1}^nZ_il_i(\thetavec; X_i).$$
Note that the term $\sum_{i=1}^nZ_i$ in \eqref{equ:sumton} is ignored because it will not affect the solution. The FRW version of the ML estimate $\thetavechats$ is the solution to $\lbars'(\thetavec)=\partial\lbars(\thetavec)/\partial\thetavec=0.$ The proofs of the following three results are given in the appendix.

\begin{result}\label{result:consistency}
The FRW ML estimator $\thetavechats$ is consistent for $\thetavec$ if $\thetavechat$ is consistent for $\thetavec$. That is if $\thetavechat\to\thetavec$ then $\thetavechats\to\thetavec$, as $n\to\infty.$
\end{result}
Note that the ML estimator $\thetavechat$ is consistent and asymptotically unbiased under some mild conditions (e.g., pages 309-310 of \citeNP{coxhinkley1974}). \textbf{Result~\ref{result:consistency}} shows that the FRW bootstrap estimator is also consistent, and thus it is also asymptotically unbiased, which is a desirable property as sometimes bootstrap methods can generate estimates that are biased.

The asymptotic normality is related to the distribution of $\sqrt{n}(\thetavechats-\thetavechat)|\Xvec_n$ (i.e., given the random sample).
\begin{result}\label{res:normal}
The distribution of $\sqrt{n}(\thetavechats-\thetavechat)$ conditional on the random sample $\Xvec_n$ goes to $\protect{\N}\left[0,I(\thetavec)^{-1}\right]$ as $n\to\infty$. That is
$\sqrt{n}(\thetavechats-\thetavechat)|\Xvec_n\to \N\left[0,I(\thetavec)^{-1}\right]$, as $n\to \infty$. Here $I(\thetavec)$ is the Fisher information matrix for $\thetavec$ based on $\Xvec_n$.
\end{result}
Note that the ML estimator $\thetavechat$ asymptotically has a $\N[\thetavec, I(\thetavec)^{-1}]$ distribution, under some mild conditions. \textbf{Result~\ref{res:normal}} shows that the distributions of $\thetavechat$ and $\thetavechat^{\ast}$ are the same when $n$ goes to $\infty$.
Thus, one can use the distribution of a function of $\thetavechat^{\ast}$ to mimic the distribution of the corresponding function of $\thetavechat$.

Under mild conditions, the ML estimates exist for the FRW samples for the log-location-scale family of distributions with right censoring. Specifically, consider data $(t_i, \delta_i), i=1,\ldots, n$ where $t_i$ is the time to event, $\delta_i$ is the censoring indicator, and $n$ is the number of data points. The parameters are denoted by $\thetavec=(\mu,\sigma)'$ where $\mu$ is the location parameter and $\sigma$ is the scale parameter. The loglikelihood can be written as $l(\thetavec)=\sum_{i=1}^{n}l_i(\thetavec),$
where $l_i(\thetavec)$ is the log likelihood contribution from observation $i$. The weighted loglikelihood is
$l^{\ast}(\thetavec)=\sum_{i=1}^{n}w_il_i(\thetavec)$ (here, the normalized weights $w_i$ are used for convenience).

\begin{result}\label{res:mle.exist}
For data with right censoring that are generated from commonly used log-location-scale family of distributions (e.g., lognormal and the Weibull), the minimum condition for the ML estimate to exist for $l^{\ast}(\thetavec)$, is either (i) two distinct failure times $t_1$ and $t_2$, or (ii) one failure time $t_1$ and a right-censored observation $t_2$ with $t_2>t_1$.
\end{result}
Because of the continuous weights (i.e., all $w_i$'s are positive), a failure will always make a contribution to the likelihood in the FRW samples. \textbf{Result~\ref{res:mle.exist}} indicates that the requirement for the existence of the ML estimate is mild.

\section{Applications to Confidence Intervals}\label{secACI}
In this section, we use three examples to illustrate the applications of FRW in the construction of confidence intervals.

\subsection{Bearing Cage Field Failure Data}
\subsubsection{Background}
There were 1703 aircraft engines that had been put into service over time, as shown in the event plot in
Figure~\ref{fig:bearing.cage.event.prob}(a). There had been 6 failures and there were 1697 right-censored observations. These data were
originally given in \shortciteN{abmr83} and were re-analyzed in Chapter~8 of \citeN{me98}.

\subsubsection{Weibull Analysis}
\begin{figure}
\begin{center}
\begin{tabular}{cc}
\includegraphics[width=.465\textwidth]{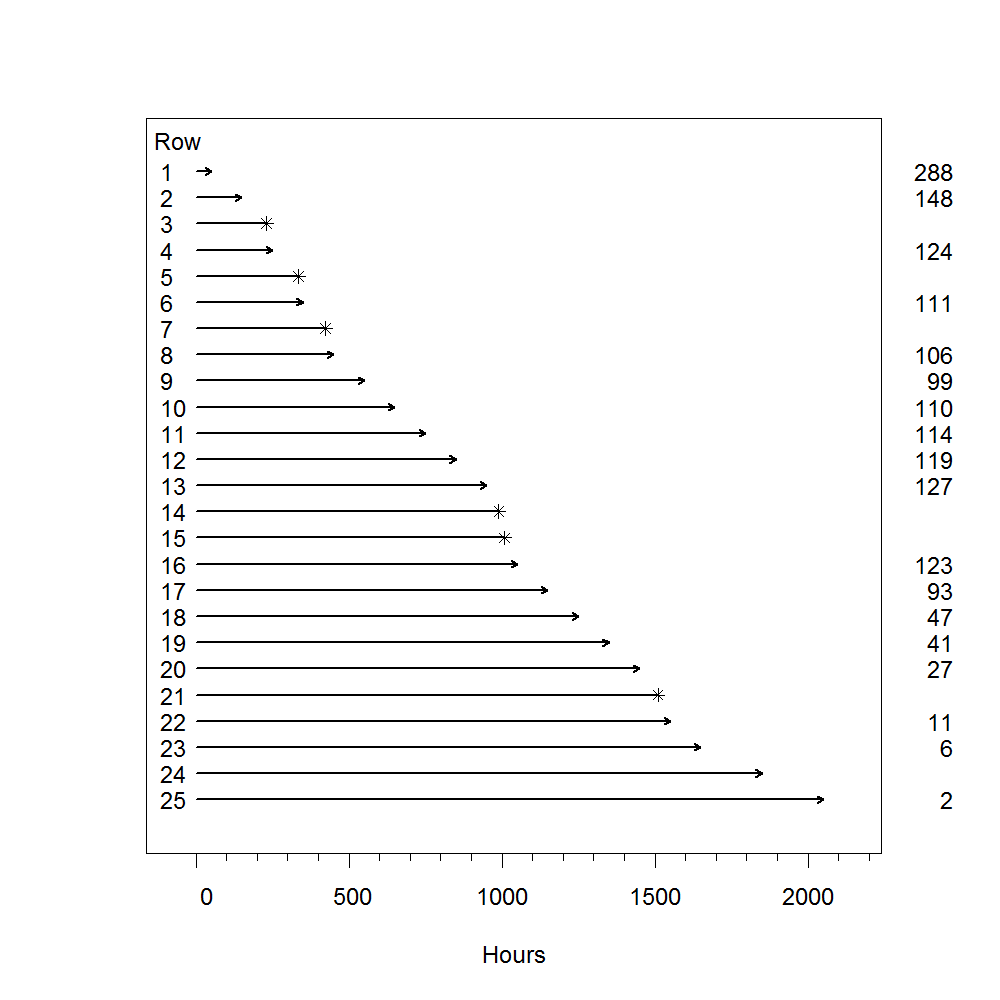}&
\includegraphics[width=.45\textwidth]{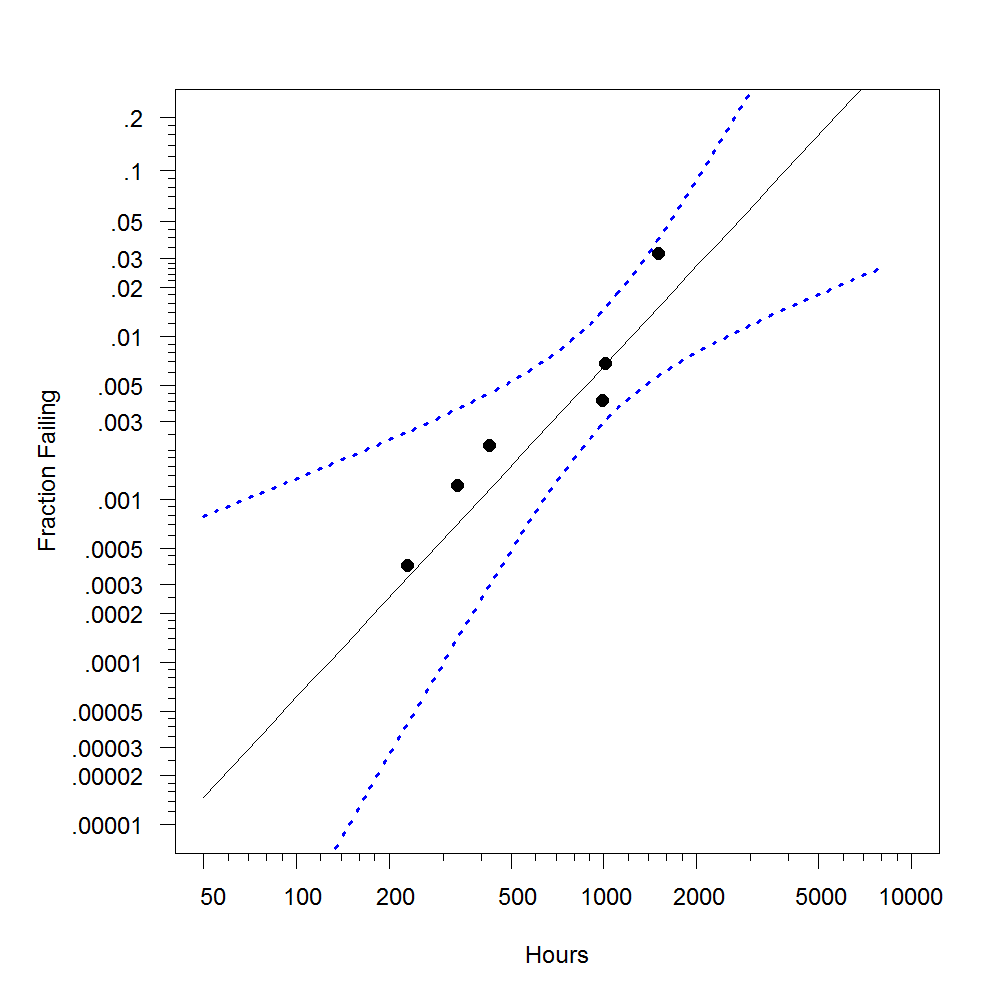}\\
(a) Event Plot & (b) Probability Plot
\end{tabular}
\end{center}
\caption{Event plot and Weibull probability plot for the bearing cage field-failure data.}\label{fig:bearing.cage.event.prob}
\end{figure}

The Weibull distribution with cdf
$$F(t;\eta,\beta)=\Pr(T\leq t)=1-\exp\left[-\left(\frac{t}{\eta}\right)^\beta\right],\;t>0,$$
will be used to describe these data, where $T$ is the time to failure, $\eta=\exp(\mu)$ is the scale parameter, and $\beta=1/\sigma$ is the shape parameter, while $\mu$ and $\sigma$ are location and scale parameters respectively for $\log(T)$. Figure~\ref{fig:bearing.cage.event.prob}(b) is a Weibull probability plot of the field-failure data. Table~\ref{tab:bearing.cage.ML} summarizes the numerical results of the estimation. For this example, we will focus on the estimation of the Weibull shape parameter $\beta$.
The ML estimate is 2.035. The upper endpoint of the Wald 95\% confidence
interval is 5.67. Because of the small number of failures, the Wald confidence interval is not trustworthy. The likelihood upper endpoint is 3.58. Another alternative for computing trustworthy confidence intervals is the
bootstrap. Care is needed, however, when using the resampling bootstrap method with heavy
censoring. If the expected number failing is too small there could be bootstrap samples with only 0 or 1 failures, possibly causing the ML algorithm to fail. As described in \textbf{Result~\ref{res:mle.exist}} of Section~\ref{secTR}, there is a unique maximum of the likelihood if there is at least one failure, as long as there is at least one censored observation greater than that failure. It is, however, possible that the maximization algorithm will fail in such cases because the shape of the likelihood can be poorly behaved. For the Bearing Cage example, the probability of obtaining a bootstrap sample with 0 or 1 failures using the resampling method is 0.017 based on a simple binomial
distribution computation. Using the FRW method, the probability is~0!

\begin{table}
\begin{center}
\caption{ML estimates for the Weibull analysis of the ball bearing life test data.}\label{tab:bearing.cage.ML}
\vspace{1.5ex}
\begin{tabular}{crrrr}\hline\hline
\multirow{2}{*}{Parameter} & \multirow{2}{*}{Estimate} & \multirow{2}{*}{Std Error} &\multicolumn{2}{c}{95\% CI}\\ \cline{4-5}
&	&	&  Lower 	& Upper \\\hline
$\eta$  & 11792.178	&9848.1267	&2294.6744	&60599.215 \\
$\beta $  & 2.035	    &0.6657	    &1.2403	    &5.670   \\\hline\hline
\end{tabular}
\end{center}
\end{table}

\subsubsection{Bootstrap Results}
\begin{figure}
	\begin{center}
		\includegraphics[width=.8\textwidth]{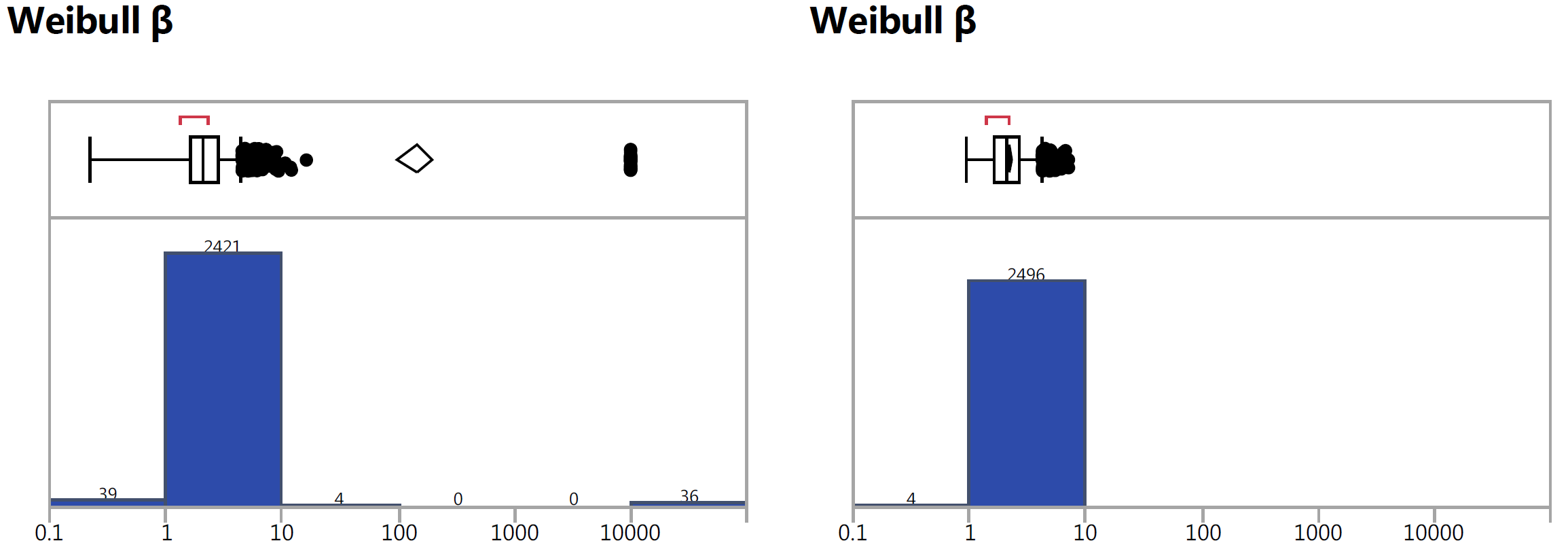}
	\end{center}
	\caption{Resampling (left) and fractional-random-weight (right) bootstrap results for the Weibull shape parameter for the bearing cage field-failure data.}\label{fig:bc.frw}
\end{figure}

Figure~\ref{fig:bc.frw} shows results from the resampling (left) and the FRW bootstrap for the Weibull shape
parameter $\beta$. The histogram on the left shows that there were 36 samples that resulted in a wild
estimate of $\beta$ which were probably caused by having resamples with 0 failures. The upper endpoint
of the confidence interval is even more extreme than that provided by the untrustworthy Wald method.
The histogram on the right, based on the FRW bootstrap method is better behaved and the upper
endpoint only 4.4. This is a more trustworthy value and is consistent with common experience with
fatigue failures in the field. Interestingly (but not surprisingly) the FRW method runs somewhat faster
than the resampling method for this example. This is because with the FRW method the optimization
algorithm is not faced with bootstrap samples that result in poorly behaved likelihoods that require a lot
of time trying to find a maximum that does not exist.

\begin{table}
\begin{center}
\caption{Resampling and FRW bias-corrected (BC) percentile bootstrap results for the Weibull shape parameter for the bearing cage field-failure data.}\label{tab:fig5}
\vspace{1.5ex}
\begin{tabular}{crr | crr}\hline\hline
	\multicolumn{3}{c|}{Resampling}  & \multicolumn{3}{c}{Fractional-Random-Weight} \\
	\multicolumn{3}{c|}{Bootstrap Confidence Limits}  & \multicolumn{3}{c}{Bootstrap Confidence Limits} \\ \hline
	Confidence Level & BC Lower & BC Upper  & Confidence Level & BC Lower & BC Upper   \\ \hline
	0.95     & 1.035  & 6.540    & 0.95     & 1.188   & 4.402 \\
	0.90      & 1.153  & 4.866    & 0.90     & 1.270  & 3.898 \\
	0.80      & 1.304  & 3.750    & 0.80      & 1.384  & 3.344 \\
	0.50      & 1.574  & 2.671    &  0.50      & 1.625  & 2.641 \\ \hline\hline
\end{tabular}
\end{center}
\end{table}

\subsection{Rocket Motor Field-Failure Weibull Analysis}
\subsubsection{Background}
\citeN{os01} present a Bayesian analysis of rocket motor field-failure data. The data were
reanalyzed in Chapters 14 and 18 of \citeN{mhe17}. There were approximately
20,000 missiles in inventory that had been manufactured over a number of years and put into the
stockpile. There had been 1,940 rockets put into flight over a period of time up to 18 years subsequent
to their manufacture. At their time of flight, 1,937 of these motors performed satisfactorily; but there
were three catastrophic launch failures. The failure probability at 20 years was of interest.

\begin{table}
\begin{center}
\caption{Rocket motor life data (in years since manufacture).}\label{tab:fig7}
\vspace{1.5ex}
\begin{tabular}{  l  c | l  c|  l  c  }
	\hline\hline
\multirow{2}{*}{Years} & Number   & \multirow{2}{*}{Years}   & Number   &  \multirow{2}{*}{Years}   & Number \\
	 & of Motors&   & of Motors&     & of Motors\\ \hline
	$> 1$ & 105      & $>  8$ & 211      & $> 14$   & 14 \\
	$> 2$ & 164      & $>  9$ & 124      & $> 15$   & 5 \\
	$> 3$ & 153      & $> 10$ & 90       &  $> 16$  & 3 \\
	$> 4$ & 236      & $> 11$ & 72       &  $< 8.5$ & 1 \\
	$> 5$ & 250      & $> 12$ & 53       &  $<14.2$ & 1 \\
	$> 6$ & 197      & $> 13$ & 30       &  $<16.5$ & 1 \\
	$> 7$ & 230      &  &  &  &  \\ \hline\hline
\end{tabular}
	\end{center}
\end{table}

\begin{figure}
\begin{center}
\begin{tabular}{cc} 	
\includegraphics[width=.465\textwidth]{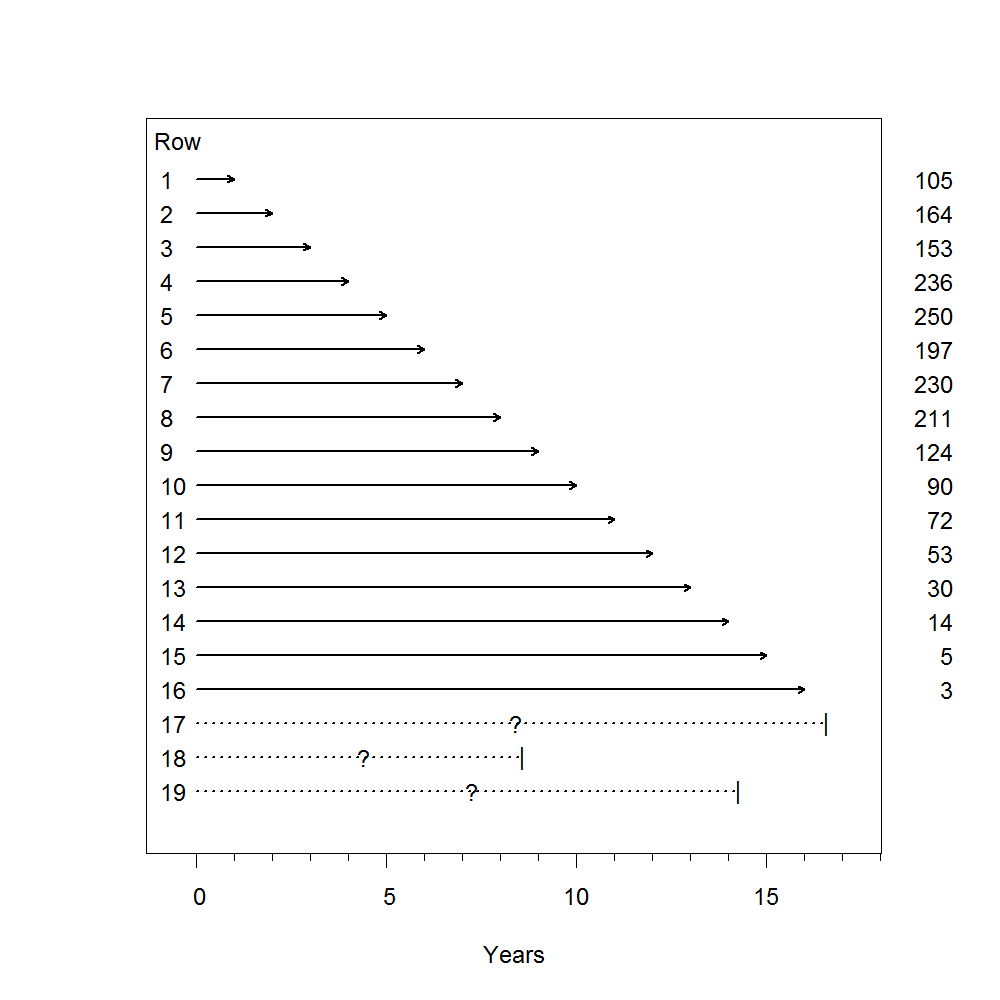} &
\includegraphics[width=.45\textwidth]{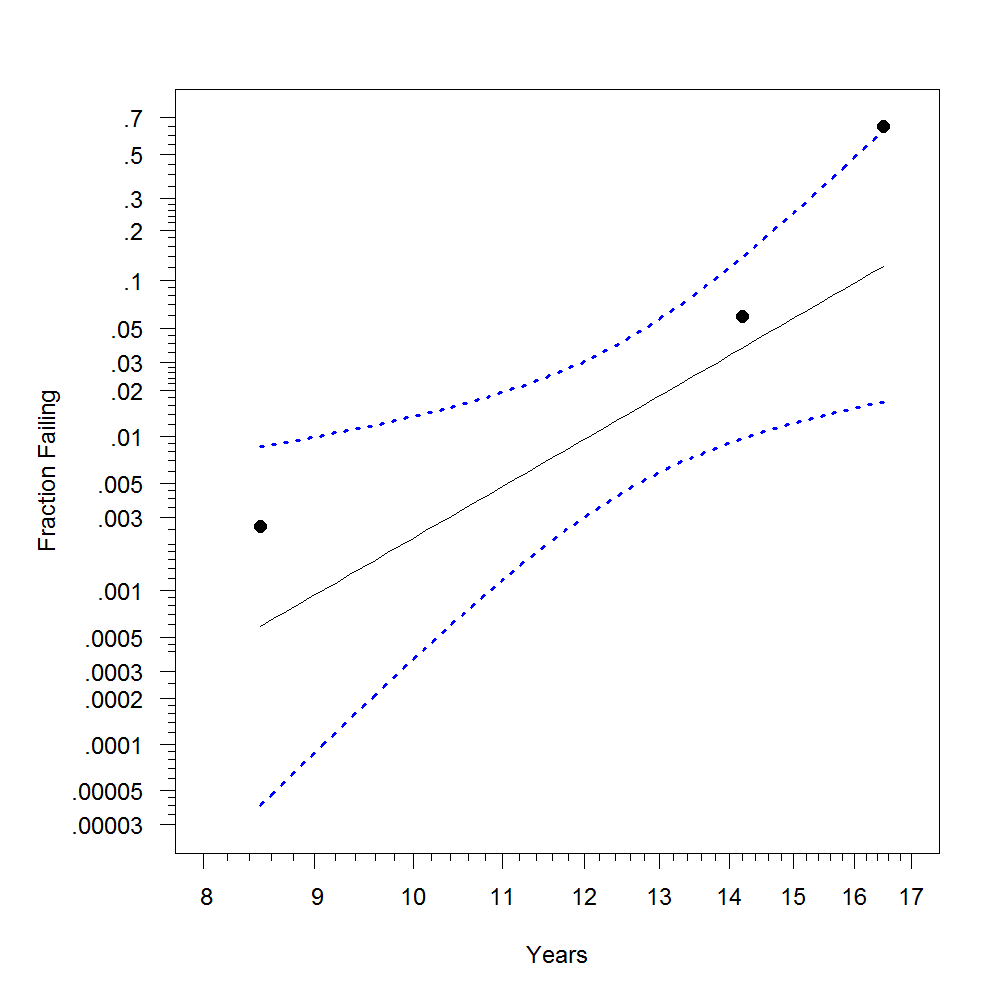}\\
(a) Event Plot & (b) Probability Plot
\end{tabular}
\end{center}
\caption{Event plot and Weibull probability plot for the rocket motor field-failure data.}\label{fig:rocket.motor}
\end{figure}

The data are shown in Table~\ref{tab:fig7} where $>$ indicates that the failure time was greater that the indicated year
(right-censored observations) and $<$ indicates that the failure time was less than the indicated year (left
censored observations). Figure~\ref{fig:rocket.motor}(a) is an event plot showing the structure of the data. Although these data
could be described with a binary regression model (failure probability as a function of years since
manufactured), there are advantages of treating such data as censored failure-time data. In particular, the
fraction failing as a function of time is constrained to be monotone increasing and we can use
probability plotting to assess whether a chosen distribution is appropriate. (although there is little information for such an assessment in this application.)
The usual resampling (integer weight) bootstrap will not work well with this data set, as demonstrated
below.

\subsubsection{Weibull Analysis}
Figure~\ref{fig:rocket.motor}(b) is a Weibull probability plot of the rocket motor field failure data. Table~\ref{tab:fig10} gives the numerical
ML results and Wald confidence intervals for the distribution parameters. As with the bearing cage
example, we will again focus on the Weibull shape parameter to compare the different confidence
intervals. The upper endpoint of the Wald confidence interval for $\beta$ is 34.6. The
likelihood interval has an upper endpoint of 15.54, considerably smaller than then that for the
Wald interval. Generally, the likelihood interval would be considered to be more trustworthy than the
Wald interval. The bootstrap interval would also be expected to provide a trustworthy interval. In this
case, however, because of the heavy censoring, the resampling bootstrap method will not work properly.

\begin{table}
\begin{center}
\caption{ML estimation results for the Weibull analysis and likelihood-based confidence intervals for the rocket motor field failure data.}\label{tab:fig10}
\vspace{1.5ex}
\begin{tabular}{crrrr}
	\hline\hline
\multirow{2}{*}{Parameter}&\multirow{2}{*}{Estimate} &\multirow{2}{*}{Std Error} &\multicolumn{2}{c}{95\% CI}\\ \cline{4-5}
	       &  &  &  Lower & Upper \\\hline
	$\eta$ & 21.228& 4.591& 16.846  & 67.396 \\
	$\beta$ & 8.126& 3.172&  2.963  & 15.541 \\
	\hline\hline
\end{tabular}
\end{center}
\end{table}

\subsubsection{Bootstrap Results}
\begin{figure}
	\begin{center}
		\includegraphics[width=.8\textwidth]{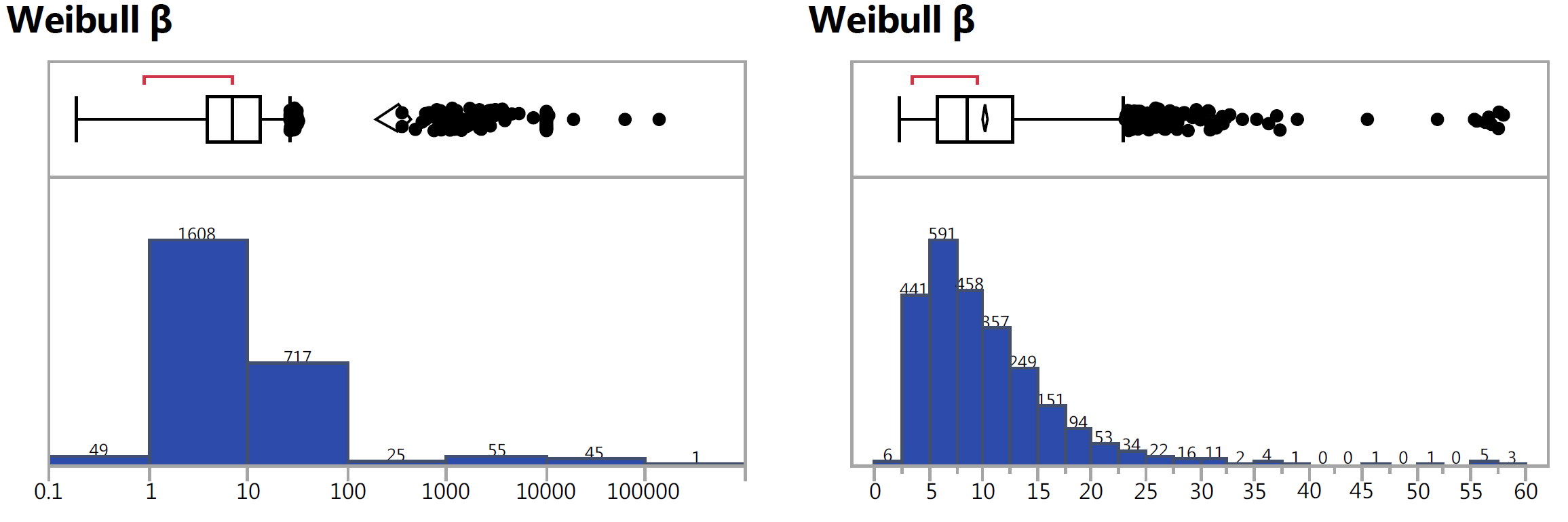}
	\end{center}
	\caption{Resampling (left) and fractional-random-weight (right) bootstrap results for the Weibull
		shape parameter for the rocket motor field-failure data.}\label{fig:fig11}
\end{figure}

Figure~\ref{fig:fig11} shows the results for bootstrapping the ML estimate of the Weibull shape parameter for the
rocket motor. Using the resampling method there was a large number of samples giving a very large
value of the bootstrap estimate of $\beta$. These probably arose from bootstrap samples that had no or only
one of the left-censored observations in the bootstrap sample. Using the FRW bootstrap, on the other
hand resulted in a much more reasonable distribution and corresponding 95\% confidence interval for $\beta$. Table~\ref{tab:fig12} shows the resampling and fractional-random-weight bootstrap results for the Weibull shape parameter for the rocket motor field-failure data.

\begin{table}
\begin{center}
\caption{Resampling and FRW bias-corrected (BC) percentile bootstrap results for the Weibull shape parameter $\beta$ for the rocket motor field-failure data.}\label{tab:fig12}
\vspace{1.5ex}
\begin{tabular}{crr | crr}\hline\hline
	\multicolumn{3}{c|}{Resampling}  & \multicolumn{3}{c}{Fractional-Random-Weight} \\
	\multicolumn{3}{c|}{Bootstrap Confidence Limits}  & \multicolumn{3}{c}{Bootstrap Confidence Limits} \\ \hline
	Confidence Level & BC Lower & BC Upper  & Confidence Level & BC Lower & BC Upper   \\ \hline
			0.95      & 1.000        & 10000.000   &0.95      & 2.832  & 22.912 \\
			0.90      & 1.210   & 2455.210 &0.90      & 3.097  & 19.305 \\
			0.80      & 2.296   & 30.227  &0.80      & 3.666  & 15.988 \\
			0.50      & 5.348   & 19.461 &0.50      & 5.207  & 11.479 \\ \hline\hline
		\end{tabular}
	\end{center}
\end{table}

\subsection{Bootstrapping the Generalized Gamma Distribution Model for the Ball Bearing Failure Time Data}
\subsubsection{Background}
\begin{figure}
	\begin{center}
\begin{tabular}{cc}
\includegraphics[width=.465\textwidth]{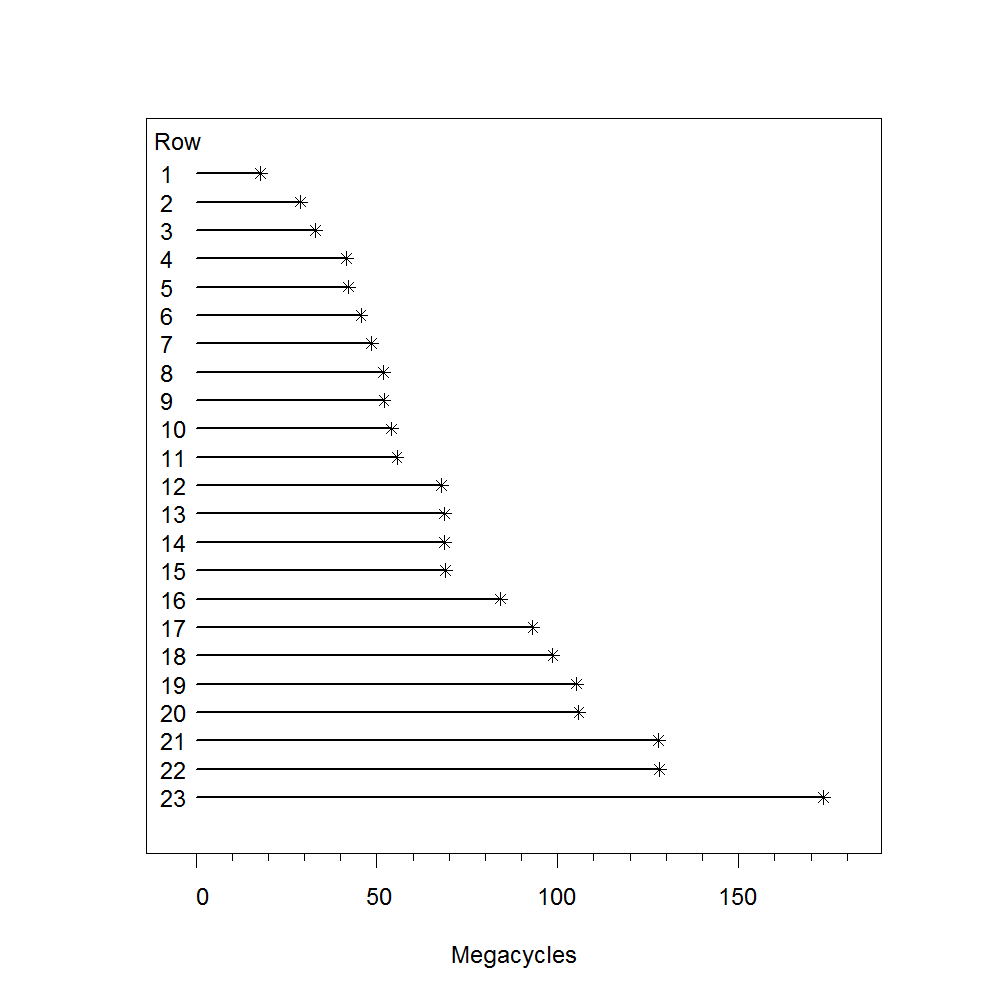}&
\includegraphics[width=.45\textwidth]{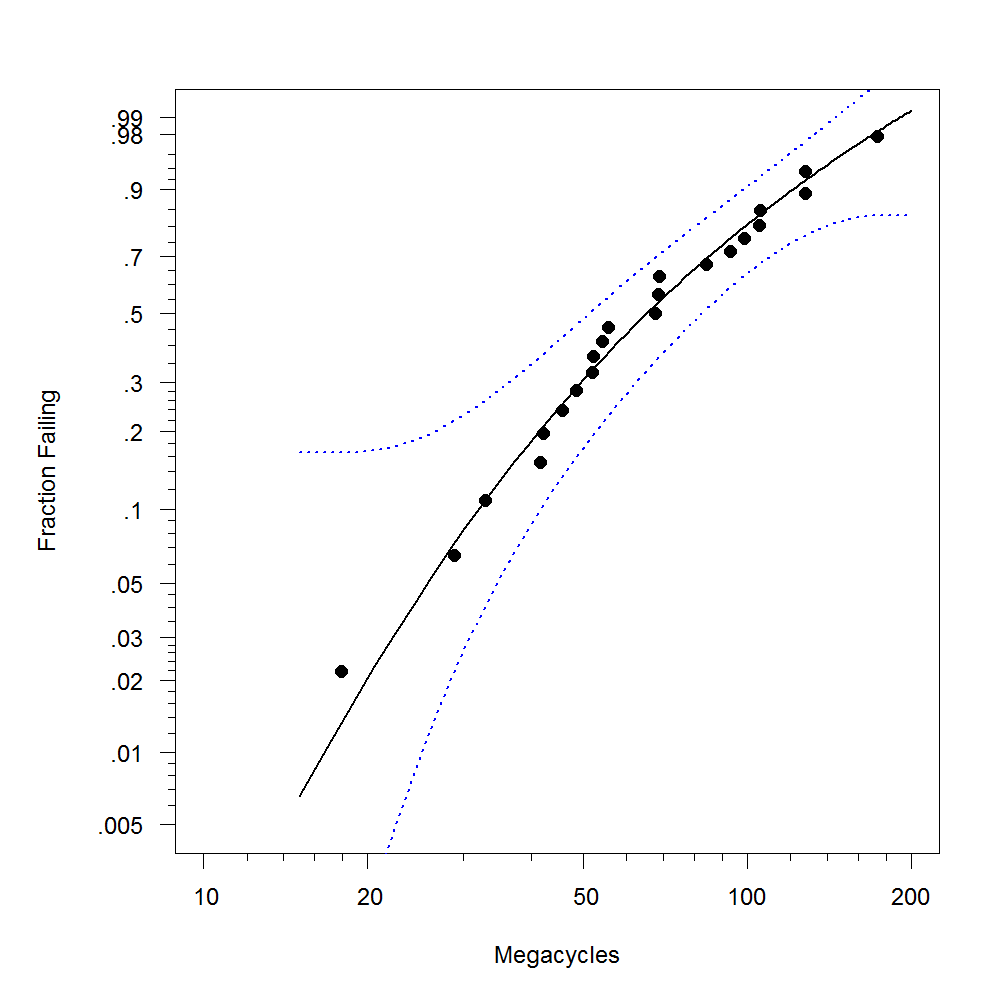}\\
(a) Event Plot & (b) Probability Plot
\end{tabular}
	\end{center}
	\caption{Event plot and Weibull probability plot for generalized gamma fit to the ball bearing life test data.}\label{fig.fig14}
\end{figure}

\citeN{me98} and \citeN{l03} fit the generalized gamma distribution to ball bearing life
test data that were originally reported in \citeN{lz56}. Figure~\ref{fig.fig14}(a) is an event plot of the
data. There was no censoring. The generalized gamma distribution is interesting in that depending on the value of the shape
parameter $\lambda$, the Weibull ($\lambda = 1$), lognormal ($\lambda = 0$), and Frechet ($\lambda=-1$) distributions are special cases. The detailed expression for the cdf of the generalized gamma distribution is as follows.
\begin{align*}
F(t;\mu,\sigma,\lambda)=
\begin{cases}
	\Phi_{\text{lg}}\left[\lambda \omega+\log\left(\lambda^{-2}\right),\lambda^{-2}\right],& \text{if } \lambda>0\\
	\Phi_{\text{nor}}(\omega),              & \text{if }\lambda=0\\
	1-\Phi_{\text{lg}}\left[\lambda \omega+\log(\lambda^{-2}),\lambda^{-2}\right],	& \text{if }\lambda<0,\\
\end{cases}
\end{align*}
where $t>0$, $\omega=\left [\log(t)-\mu \right ]/\sigma$, $\mu$ is a location parameter, $\sigma$ is a scale parameter, and $\lambda$ is a shape parameter. Here $\Phi_\textrm{lg}(z,\kappa)=\Gamma_{\textrm{I}}[\exp(z),\kappa]$, $\phi_\textrm{lg}(z,\kappa)=\exp \left [{\kappa z-\exp( z)} \right]/\Gamma(\kappa)$, $\Gamma_{\textrm{I}}(v,\kappa)=\int^{v}_{0} x^{\kappa-1}\exp(-x)dx/\Gamma(\kappa)$ is the incomplete gamma function, and $\Gamma(z)$ is the gamma function.

\subsubsection{Maximum Likelihood Estimation}
\begin{figure}
	\begin{center}
		\includegraphics[width=.5\textwidth]{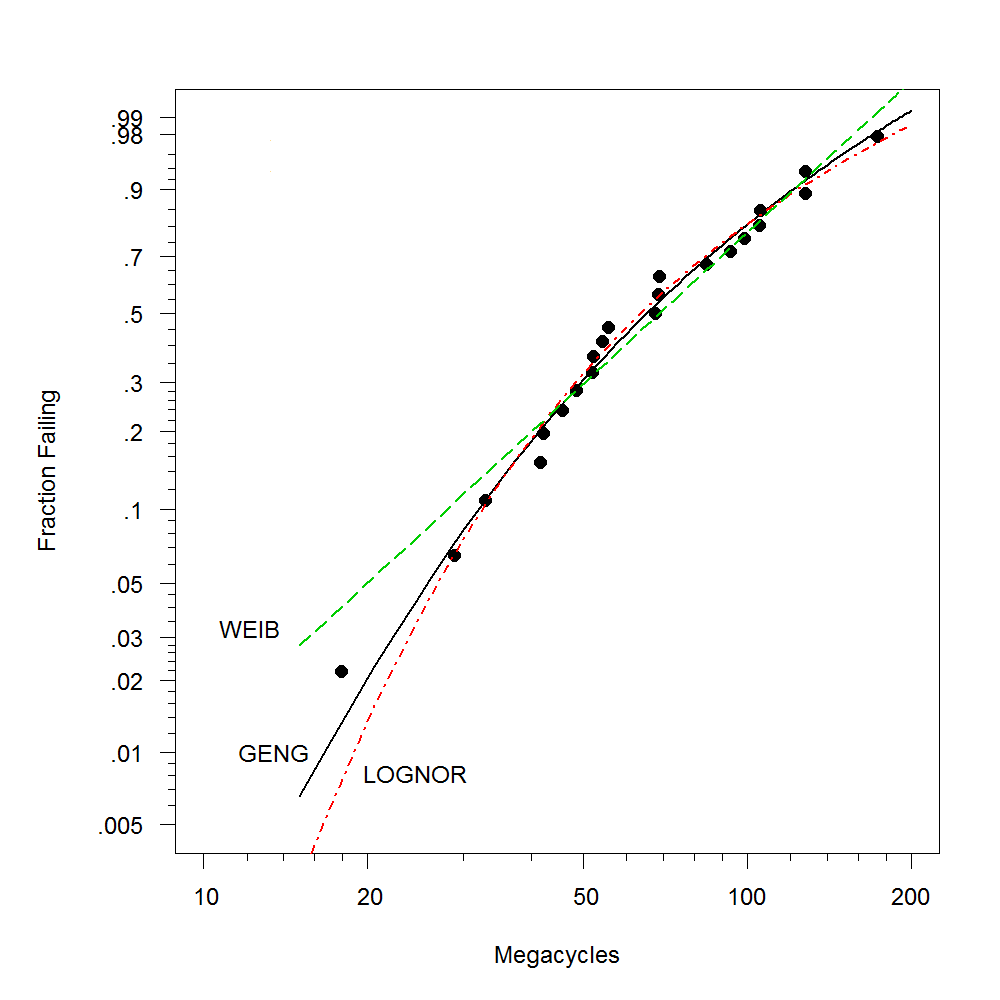}
	\end{center}
	\caption{Ball bearing failure time data fitted by the Weibull, generalized gamma, and lognormal distributions on the Weibull probability paper.}\label{fig:fig20}
\end{figure}

Figure~\ref{fig.fig14}(b) shows the ML estimates of the cdf of the generalized gamma distribution, plotted on Weibull probability paper.
We can see that it generally fits the data well.  Figure~\ref{fig:fig20} shows ML estimates of the Weibull, generalized gamma, and lognormal cdfs on Weibull probability paper. Table~\ref{tab:fig16} gives the ML estimates and Wald
confidence intervals for the parameters. The 95\% likelihood-based confidence interval for $\lambda$ is $\left[-0.76,
1.53\right]$, somewhat wider than the Wald confidence interval. Again, the likelihood-based interval is generally
more trustworthy. In either case, the confidence interval for $\lambda$ provides an indication that both the
lognormal and Weibull distributions are consistent with the data. This is because both 0 and 1 lie inside
the confidence interval.

The bootstrap provides another alternative to computing confidence intervals for this distribution. In
the next section, we will compare the resampling and the FRW bootstrap methods.

\begin{table}
\begin{center}
\caption{ML estimation results based on generalized gamma distribution and Wald confidence intervals for the ball bearing failure time data.}\label{tab:fig16}
\vspace{1.5ex}
\begin{tabular}{  c  r  r  r  r  }
	\hline\hline
\multirow{2}{*}{Parameter} &\multirow{2}{*}{Estimate} &\multirow{2}{*}{Std Error} &\multicolumn{2}{c}{95\% CI}\\ \cline{4-5}
	 &  &  & Lower  & Upper  \\ \hline
	$\mu$        & 4.230 & 0.177 & 3.883   & 4.577 \\
	$\sigma$     & 0.510 & 0.079 & 0.354   & 0.665 \\
	$\lambda$    & 0.308 & 0.549 & $-$0.760   & 1.383 \\ \hline\hline
\end{tabular}
\end{center}
\end{table}

\subsubsection{Bootstrap Results}
\begin{figure}
	\begin{center}
		\includegraphics[width=.8\textwidth]{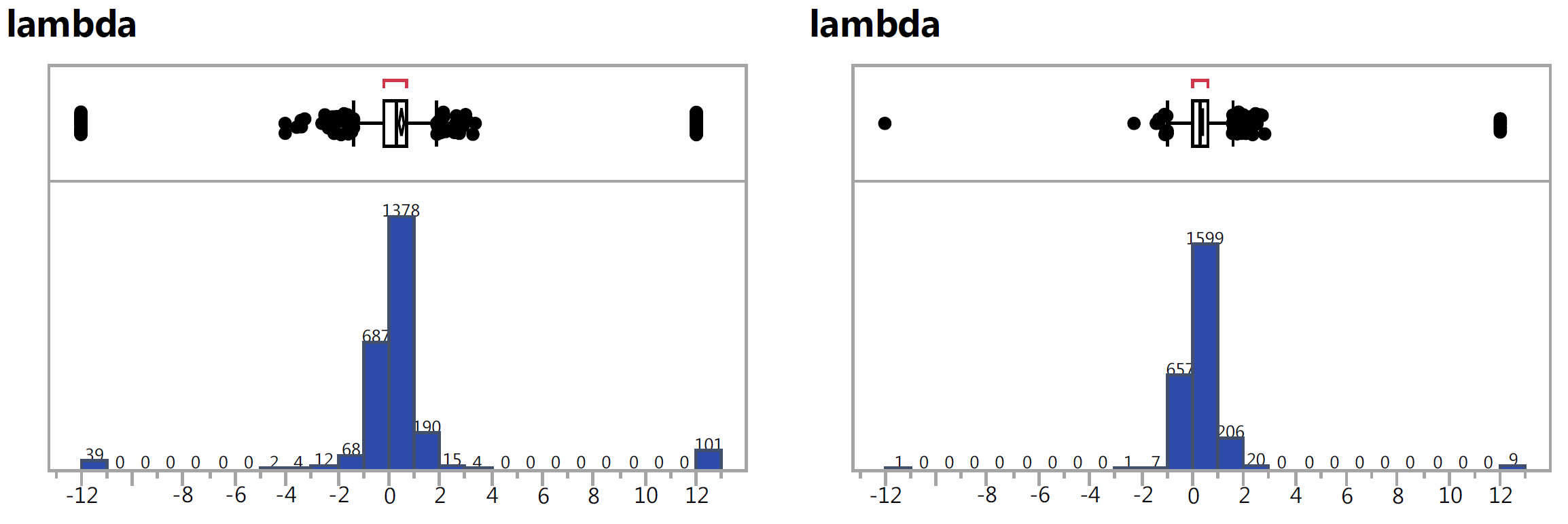}
	\end{center}
	\caption{Resampling (left) and fractional-random-weight (right) bootstrap results for
		the generalized gamma distribution shape parameter for the ball bearing data.}\label{fig:fig17}
\end{figure}

Figure~\ref{fig:fig17}  gives bootstrap results for the ball bearing generalized gamma distribution shape parameter $\lambda$. Table~\ref{tab:fig18} shows the resampling and fractional-random-weight bootstrap results for the generalized gamma distribution shape parameter. The histogram on the left of Figure~\ref{fig:fig17} shows results for the resampling bootstrap method; the histogram on the
right shows results using the FRW bootstrap method. The ML algorithm used here constrains $\lambda$ to be in the interval $\left[-12,12\right]$. With the resampling method, there was ML
estimate convergence problems with a substantial number of the bootstrap samples. In particular, there were 39 times the endpoint of the 95\% bootstrap interval of the $\lambda$ parameter was $-$12 and 101 times where the upper endpoint was +12. The $101$ values at $+12$ were more than enough to cause the upper endpoint of the $95\%$ bootstrap confidence interval to be $+12$. The corresponding numbers when using the FRW bootstrap
method were 1 and 9 times, respectively and these numbers are so small that they have little or no effect on the bootstrap confidence intervals for  $\lambda$. The FRW method provides a better method for computing confidence intervals for the generalized gamma distribution when the sample size is not large.

\begin{table}
	\begin{center}
		\caption{Resampling and FRW bias-corrected (BC) percentile bootstrap results for
			the generalized gamma distribution shape parameter for the ball bearing data.}\label{tab:fig18}
\vspace{1.5ex}
\begin{tabular}{crr | crr}\hline\hline
	\multicolumn{3}{c|}{Resampling}  & \multicolumn{3}{c}{Fractional-Random-Weight} \\
	\multicolumn{3}{c|}{Bootstrap Confidence Limits}  & \multicolumn{3}{c}{Bootstrap Confidence Limits} \\ \hline
	Coverage & BC Lower & BC Upper  & Coverage & BC Lower & BC Upper   \\ \hline
			0.95      & $-$1.318   & 12.000       &0.95      & $-$0.595  & 1.704 \\
			0.90      & $-$0.877  & 12.000         &0.90      & $-$0.433  & 1.346 \\
			0.80      & $-$0.539  & 1.243    &0.80      & $-$0.263  & 1.039 \\
			0.50      & $-$0.072  & 0.729    &0.50      & 0.0106  & 0.642 \\ \hline\hline
		\end{tabular}
	\end{center}
\end{table}

\section{An Application to Prediction Intervals}\label{secAPI}
In this section, we illustrate the use of FRW in the construction of prediction intervals.

\subsection{Background}
\begin{figure}
\begin{center}
\includegraphics[width=.55\textwidth]{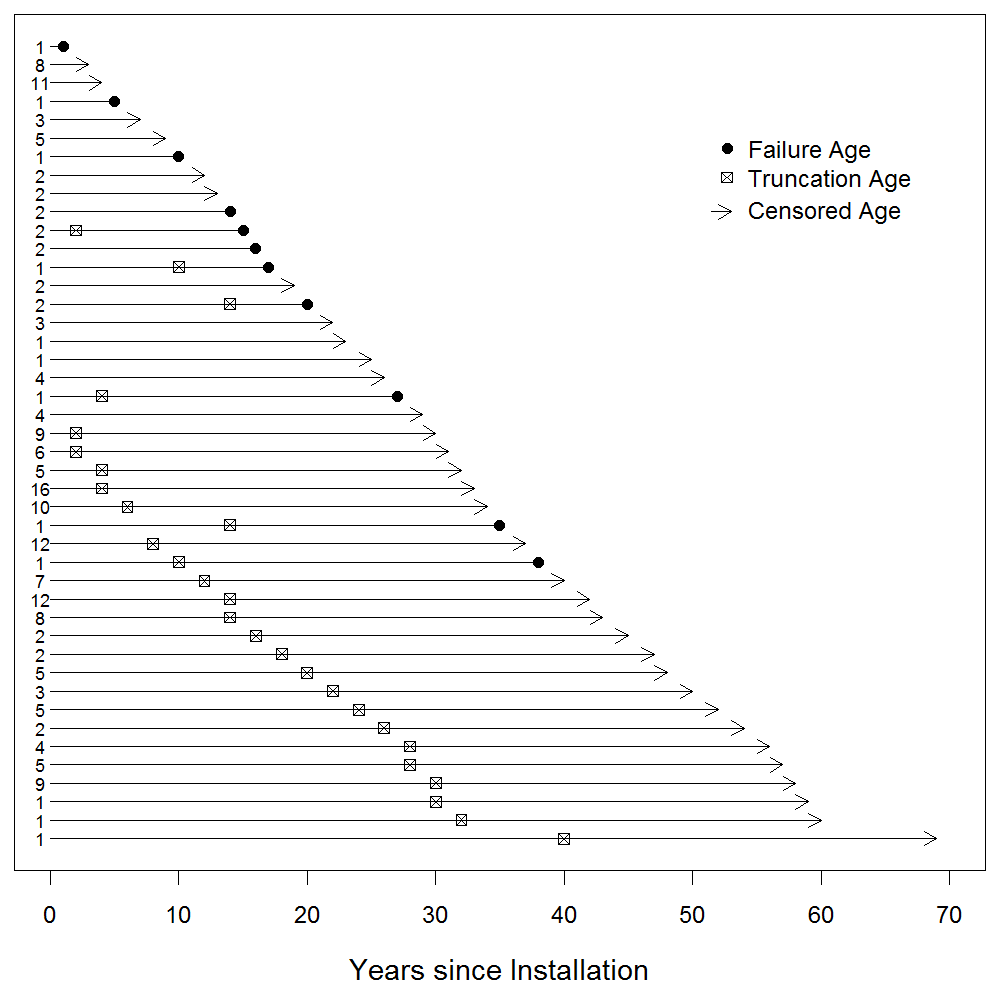}
\caption{Event plot for the power transformer field-failure data.}\label{fig:fig8}
\end{center}
\end{figure}

Extending previous work of \citeN{em99} and \citeN{lf05}, \citeN{hmm09} describe the use of the FRW bootstrap to generate prediction intervals for the number of power transformers that will need to be replaced in future years. The dataset contained information on 710 power transformers with 62 units that had failed. Units still in service at the data freeze date in March 2008 are right censored. Some units that were still in service were more than 60 years old. One difficulty with the data is that records of transformers removed from service before 1980
were not available. Thus units that had been installed before 1980 and which were still in service are
observations from a truncated distribution. Figure~\ref{fig:fig8} is an event plot of a representative subset of the data.

There were several categorical covariates, including manufacturer and cooling method that had an effect on the life
distribution. It was also learned that, even after adjustment for the other covariates, that there was an
important difference between the failure-time distributions of transformers manufactured before and
after the mid-1980s. Transformers manufactured before the mid-1980s tend to have longer lifetimes,
due to over-engineering that was practiced then.

\subsection{Modeling and Maximum Likelihood Estimation}
The Weibull and lognormal distributions were fit to the data using maximum likelihood. Stratification was
based on whether unites were manufactured before or after 1987. The likelihood function is
\begin{eqnarray*}
\label{eqn:lik}
L(\thetavec|\textrm{DATA})&=&\prod_{i=1}^{n}f(t_{i};\thetavec)^{\delta_{i}\nu_{i}}\times
\left[\frac{f(t_{i};\thetavec)}{1-F(\tau_{i}^L;\thetavec)}\right]^{\delta_{i}(1-\nu_{i})}\\\nonumber
&&\times\left[1-F(t_{i};\thetavec)\right]^{(1-\delta_{i})\nu_{i}}
\times\left[\frac{1-F(t_{i};\thetavec)}{1-F(\tau_{i}^L;\thetavec)}\right]^{(1-\delta_{i})(1-\nu_{i})}.
\end{eqnarray*}
where $t_{i}$ is the failure or censoring time, $\tau_{i}^L$ is the lower truncation time, and $\delta_{i}$ and $\nu_{i}$ are
censoring and truncation indicators respectively for transformer $i$. We use $\thetavec$ to represent the vector of parameters, and $f(t;\thetavec)$ and $F(t;\thetavec)$ are the pdf and cdf of the Weibull distribution, respectively. An important question was how to generate bootstrap samples to do the calibration of the needed prediction
intervals. The commonly-used parametric bootstrap would be complicated to implement
because it would require a model for the censoring and truncation processes. The resampling
method would also have difficulties because of the categorical covariates and the small number
of failures in some of the categories. The FRW bootstrap offered an attractive, easy-to-implement alternative that worked without any problems.

\subsection{Prediction Results}
\begin{figure}
\begin{center}
\includegraphics[width=.5\textwidth]{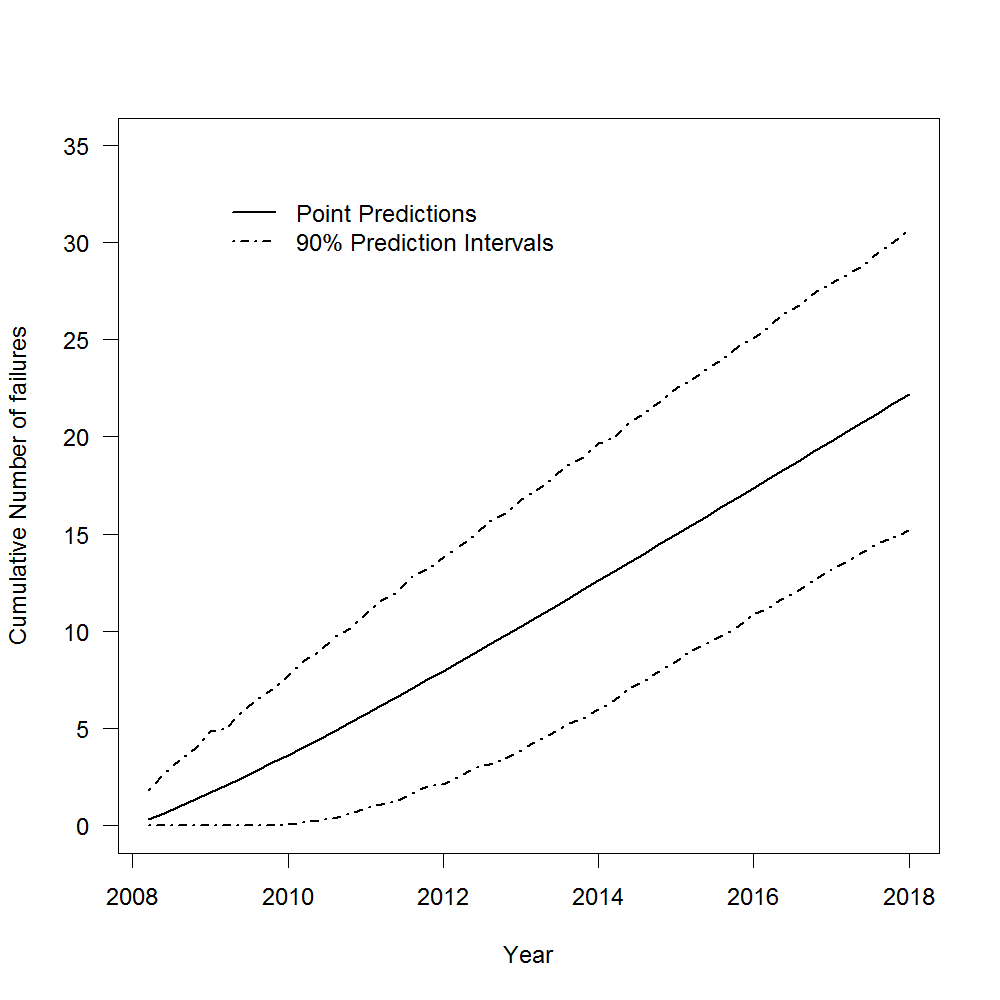}
\end{center}
\caption{Power transformer fleet predictions based on the fractional-random-weight bootstrap.}\label{fig.fig9}
\end{figure}

As an illustration, Figure~\ref{fig.fig9} shows predictions and prediction intervals for the cumulative number of transformer failures
for the next ten years, starting in 2008. The prediction is made for transformers that were installed after 1987. The predicted number
failing for the latter group is larger both because the lifetimes of the newer models tend to be shorter
and also because the size of the risk set was larger.

\begin{figure}
\begin{center}
\includegraphics[width=.55\textwidth]{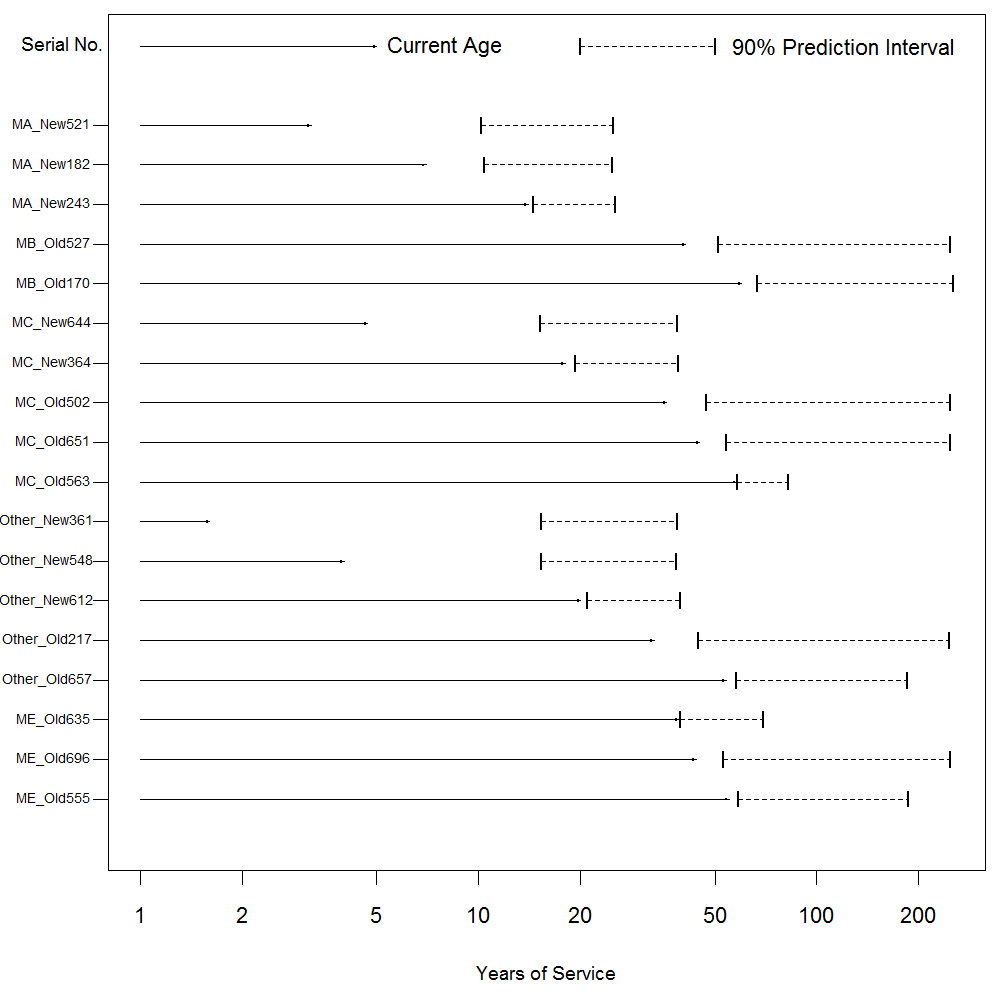}
\end{center}
\caption{Power transformer individual predictions based on the fractional-random-weight bootstrap.}\label{fig.fig10}
\end{figure}

Figure~\ref{fig.fig10} shows, for a subset of the transformers that were still in operation at the time the predictions
were made, the age of the transformer and a prediction interval quantifying the information available
about the distribution of remaining life for individual transformers. Although some of the upper
endpoints of the prediction intervals are likely overly optimistic (probably because they rely on extrapolation), the lower endpoints allowed a useful ranking of which transformers where at highest risk for failure in the short term.

\section{An Application in Design of Experiments}\label{secADE}

\subsection{Background}
Design of experiments is a common approach to problem-solving in science and industry. Designed experiments are specially structured to obtain as much information in as few samples as possible. They are usually so efficient that the removal of even small numbers of observations can induce model singularities that fundamentally change the meaning of the estimated parameters in unpredictable ways. For this reason, the resampling bootstrap is generally avoided in the analysis of designed experiments. An often-stated goal is to
obtain as much information as possible about the relationship between the experimental factors ($x$)
and the response variable(s) ($y$). Usually, a designed experiment uses a specially constructed combination of $x$ values that
optimize information gained in a small number of runs. After the data become available, then there is a
need to decide on the appropriate statistical model to describe the relationship between $x$ and $y$.

\subsection{Using the Bootstrap in Model Selection}
The bootstrap is a useful tool for identifying the subset of the $x$ variables (as well as possible interaction
and quadratic effects) that best explain variation in $y$. The resampling bootstrap, however, can
encounter problems because the removal of observations can drastically change which parameters can be estimated.
There are two well-known alternatives to resampling:

\begin{inparaitem}
\item Using a parametric bootstrap (simulating data from a given model), and

\item Resampling residuals from a fitted model.
\\
\end{inparaitem}
The problem with these two methods
is that they require specifying a model, which is what we are trying to determine! As mentioned in
Section~\ref{sec:fwboot}, the FRW bootstrap can keep all observations during the modeling process, and thus suitable for model-building applications that can be used with data from a designed experiment.

\subsection{The Nitrogen Oxides Example}
Nitrogen Oxides (NOx) are toxic greenhouse gases that are common by-products of burning organic
compounds. An experiment was done on an industrial burner to study the amount of NOx it created. A
32 run I-Optimal response surface model (RSM) design was created with 7 continuous factors:

\begin{inparaitem}
\item Hydrogen Fraction in primary fuel

\item Air/Fuel Ratio

\item Lance Position $X$

\item Lance Position $Y$

\item Secondary Fuel Fraction

\item Dispersant

\item Ethanol Percentage in primary fuel
\end{inparaitem}
\\
This design would allow estimation of all main effects, two-factor interactions and quadratic effects. We want to assess the importance of the input variables (including the two-factor interactions and quadratic terms).

\subsection{Using Forward Selection}
\begin{figure}
	\begin{center}
		\includegraphics[width=.8\textwidth]{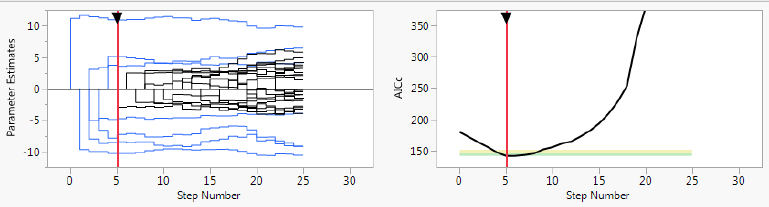}
	\end{center}
	\caption{Results for model selection using forward stepwise selection.}\label{fig:fig21}
\end{figure}

First, we apply a forward stepwise procedure that selects a model using the AIC model selection criterion. The results are shown in Figure~\ref{fig:fig21} and Table~\ref{tab:fig22}. To better understand the stability of this model choice and to explore the possibility that other variables might make an important contribution, it is possible to apply the FRW bootstrap method to the model building procedure. Then the results of such a bootstrap can be used to obtain selection probabilities for the different model terms.

\begin{table}
\begin{center}
\caption{Results from using forward stepwise selection to choose a model, showing the parameter estimates for the original predictors.}\label{tab:fig22}
\vspace{1.5ex}
\begin{tabular}{  l rrrrrr  }\hline\hline
\multirow{2}{*}{Term}   & \multirow{2}{*}{Estimate}         &  Std         &  Wald      &  Prob $>$&       \multicolumn{2}{c}{95\% CI}     \\ \cline{6-7}
 &  & Error &  ChiSquare & ChiSquare &  Lower &  Upper \\ \hline
Intercept                     & 30.305  & 0.431  & 4939.196   &   $<$0.0001 & 29.459   & 31.150 \\
Hydrogen Fraction$(0,6)$        & 2.454   & 0.341 &    51.715  &  $<$0.0001   & 1.785   &  3.123 \\
Air/Fuel Ratio$(1.05,1.3)$       & $-$2.3030   & 0.344 &    44.855 &  $<$0.0001    & $-$2.977   & $-$1.629 \\
Lance Position $X(21,25)$       & 0.853   & 0.305 &    7.800 &  $<$0.0052     & 0.254 &  1.451 \\
Lance Position $V(5,10)$        & 0 & 0 & 0 & 1 & 0 & 0 \\
Sec. Fuel Fraction$(0,6)$  & $-$1.101  & 0.261  & 17.817 & $<$0.0001 & $-$1.612 & $-$0.590 \\
Dispersant$(0.5,1.5)$ & 0 & 0 & 0 & 1 & 0 & 0 \\
Ethanol$(0,10)$ & 0 & 0 & 0 & 1.000 & 0 & 0 \\
Hydrogen Fraction & 0 & 0 & 0 & 1 & 0 & 0 \\\hline\hline
\end{tabular}
\end{center}
\end{table}

\subsection{Bootstrapping the Forward Selection Procedure}
We use the FRW approach to bootstrap the forward selection procedure. One thousand FRW bootstrap data sets were
generated. For each FRW bootstrap data set, the forward selection procedure is applied and the
corresponding row in the table gives the values of the regression coefficients. The zeros in the table
indicate that the variable was not included in the model for that bootstrap sample.

\begin{figure}
\begin{center}
\includegraphics[width=1.\textwidth]{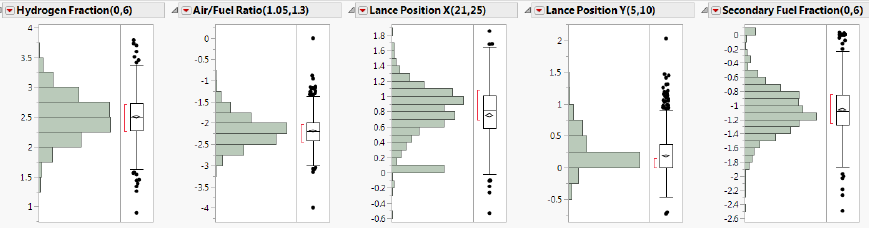}
\end{center}
\caption{Histograms that summarize the FRW Bootstrap modeling results.}\label{fig:fig24}
\end{figure}

Table~\ref{tab:fig23} shows partial results for the first 16 FRW bootstrap data sets.(i.e., for some of the regression coefficients).  Figure~\ref{fig:fig24} shows the histograms that summarize the FRW bootstrap modeling results. The spikes at 0 in some of the histograms indicate the
number of times that the corresponding variable did not enter the model (frequently for Lance Position
$Y(5,10)$ and never for Hydrogen Fraction$(0,6)$).

The results in Table~\ref{tab:fig23} can be used to construct a table like that in Table~\ref{tab:fig25}. This table gives the
proportion of times across the 1000 bootstrap samples that each variable was chosen to be in the
model. Because of the Bayesian bootstrap result given in \citeN{r81}, these proportions can be
interpreted as the posterior probability that the corresponding model term should be in the model. One
could then use a cutoff point (such as 0.50) to decide whether model terms should be included or not.

\begin{table}
\begin{center}
\small
\caption{Table showing a snapshot of the bootstrap estimates for the first 16 FRW bootstrap data sets (only the estimates of the first five variables are shown). }\label{tab:fig23}
\vspace{1ex}
\begin{tabular}{rrrrr}
	\hline\hline
	Hydrogen Fraction	& Air/Fuel Ratio	& Lance Position $X$ &	Lance Position $Y$	&Sec. Fuel Fraction\\
	$(0, 6)$	&$(1.05, 1.3)$	& $(21, 25)$ &	$(5, 10)$	&$(0, 6)$\\\hline
	2.450& $-$2.303&	0.853	&0	       &  $-$1.101 \\
	2.231& $-$2.196&	0.964	&0	       &  $-$1.137 \\
	3.006& $-$2.242&	0.714	&0	       &  $-$1.306 \\
	2.657& $-$1.929&	0.915	& 0.235   &  $-$1.154 \\
	2.987& $-$2.546&	0.085	& 0.459   &  $-$1.528 \\
	2.163& $-$1.567&	0.945	&$-$0.067   &  $-$0.761 \\
	1.804& $-$2.007&	1.113	&0	       &  $-$1.629 \\
	2.830& $-$2.549&	1.061	&0.259    &  $-$1.198 \\
	2.427& $-$2.035&	1.004	&0	       &  $-$1.497 \\
	2.487& $-$1.881&	0.345	&0	       &  $-$0.736 \\
	2.617& $-$2.145&	0	    &0.601	   &  $-$0.731 \\
	1.762& $-$1.098&	1.380	&0	       &  $-$1.446 \\
	2.416& $-$2.244&	1.075	&0.518    &  $-$1.105 \\
	2.697& $-$2.058&	1.064	&0.570    &  $-$0.890 \\
	3.028& $-$2.179&	1.190	&0.252    &  $-$1.121 \\
	2.649& $-$2.193&	0	      &0	   &  $-$1.488 \\\hline\hline
\end{tabular}
\end{center}
\end{table}

\begin{table}
\begin{center}
\caption{The proportion of times across the 1000 FRW bootstrap samples that each variable was chosen to be in the model.}\label{tab:fig25}
\vspace{2ex}
\begin{tabular}{  l  c}
	\hline\hline
	Term & Proportion Selected \\ \hline
	Hydrogen Fraction$(0, 6)$ & 1.000 \\
	Air/Fuel Ratio$(1.05, 1.3)$ & 0.999 \\
	Secondary Fuel Fraction$(0,6)$  & 0.981 \\
	Air/Fuel Ratio $\ast$ Air/Fuel Ratio  & 0.946 \\
	Lance Position $X(21, 25)$ & 0.903 \\
	Lance Position X $\ast$ Secondary Fuel Fraction  & 0.848 \\
	Hydrogen Fraction $\ast$ Secondary Fuel Fraction  & 0.733 \\
	Lance Position $Y(5, 10)$ & 0.593 \\
	Dispersant$(0.5, 1.5)$ & 0.551 \\
	Secondary Fuel Fraction $\ast$ Secondary Fuel Fraction  & 0.479 \\
	Lance Position $Y$ $\ast$ Lance Position $Y$ & 0.294 \\
	Hydrogen Fraction $\ast$ Lance Position $Y$ & 0.224 \\
	Hydrogen Fraction $\ast$ Hydrogen Fraction & 0.203 \\
	Lance Position $Y$ $\ast$ Dispersant  & 0.192 \\
	Hydrogen Fraction $\ast$ Lance Position $X$ & 0.180 \\
	Air/Fuel Ratio $\ast$ Lance Position $Y$  & 0.177 \\
	Air/Fuel Ratio $\ast$ Dispersant  & 0.163 \\
	Ethanol$(0, 10)$ & 0.113 \\
	Hydrogen Fraction $\ast$ Air/Fuel Ratio  & 0.104 \\
	Lance Position $X$ $\ast$ Lance Position $X$ & 0.083 \\ \hline\hline
\end{tabular} \end{center}
\end{table}

\section{Concluding Remarks and Areas for Future Research}\label{secCRAFR}
With vastly improved computing capabilities and the bootstrap theory that has been developed over the
past 40 years, bootstrapping provides an important useful tool for obtaining confidence intervals, prediction intervals, and better regression models.

The FRW bootstrap tremendously expands the potential areas of application of the bootstrap to
applications involving heavy censoring and/or truncation, categorical explanatory variable, and
designed experiments where dropping certain combinations of the original observations can cause
estimability problems. Those problems do not arise when the FRW bootstrap is used.

Overall, we observe that the FRW bootstrap is as easy to implement as the resampling bootstrap, and it has similar desirable properties as the resampling bootstrap in situations where the resampling bootstrap works well. The FRW bootstrap also retains desirable properties even when the resampling bootstrap breaks down. Through the examples in this paper, we see that the FRW bootstrap as a safer, more broadly applicable, alternative to the resampling bootstrap.

There are a number of areas that could be investigated to provide further insight into when and how the
FRW bootstrap methods should be used with finite samples.

\begin{inparaitem}
\item There are different, asymptotically equivalent ways to choose the random
weighs for bootstrapping (including resampling). This leaves open the question about differences in the properties of bootstrap procedures in finite samples. For example, if weights are chosen to have a mean and variance of one, what would be the effect on the performance
of varying the third or higher moments?

\item We have demonstrated a clear advantage for the FRW bootstrap in situations where estimability
problems occur when certain combinations of observations are dropped. In situations where
there will be no estimability problems is it possible that the FRW approach has other advantages. It would be useful to
compare different nonparametric and parametric methods for generating bootstrap estimates
when using a parametric model to describe one's data. In particular, it would be interesting to
compare, resampling methods, a fully parametric bootstrap simulation (e.g., where the censoring distribution is
modeled), and FRW bootstrap to see if there are important differences in bootstrap performance.

\item Generalized fiducial inference (GFI), which is also known as generalized pivotal quantity (see
\citeNP{hip06}, \citeNP{MajumderHannig2016}, and \shortciteNP{hanningetal2016}), has proven to be a powerful tool for defining confidence
interval procedures for non-standard models. Implementing GFI methods generally requires
computing a large set of simulated parameter estimates, in a manner that is similar to the
parametric bootstrap. In situations involving heavy censoring, even the parametric bootstrap
sampling will have estimability problems. Use of FRW instead should allow GFI methods to be
used in such applications.
\end{inparaitem}

\section*{Acknowledgments}
The authors thank Dan Nordman from Iowa State University for his helpful comments on an earlier version of the paper.

\appendix
\section{Proof of Theoretical Results}\label{sec:theoretical.results}
In this appendix, we prove the theoretical results given in Section~\ref{secTR}.

\subsection{Result~\ref{result:consistency}: Consistency}
Let $h(\thetavec)=\E[l_i(\thetavec; X_i)]$, $i=1,\ldots, n$. Note that $\E[\lbar(\thetavec)]=h(\thetavec)$. Also, $\V[\lbar(\thetavec)]\to0$, as $n\to\infty$, because $\thetavechat$ is consistent. The expectation of $\lbars(\thetavec)$ is
\begin{align}\label{eq1}
\E\left[\lbars(\thetavec)\right]&=\frac{1}{n}\sum_{i=1}^n\E\left[Z_il_i(\thetavec; X_i)\right]=
\frac{1}{n}\sum_{i=1}^n\E\left\{\E\left[Z_il_i(\thetavec; X_i)\right]|X_i\right\}\\\nonumber
&=\frac{1}{n}\sum_{i=1}^n\E\left[l_i(\thetavec;X_i)\right]=h(\thetavec).
\end{align}
In \eqref{eq1}, two layers of expectation are involved, in which the inner expectation is taken with respect to $Z_i$ given $X_i$ and the outer expectation is taken with respect to $X_i$. In addition,
\begin{align}\label{eq2}
\V\left[\lbars(\thetavec)\right]&=\V\left\{\E\left[\lbars(\thetavec)|\Xvec\right]\right\}+\E\left\{\V[\lbars(\thetavec)|\Xvec]\right\}\\\nonumber
&=\frac{1}{n}[2\V[l_i(\thetavec, X_i)]+h(\thetavec)^2]\to0,
\end{align}
as $n\to\infty$. Similarly, for the variance/expectation pairs in \eqref{eq2}, the inside operator is taken with respect to $Z_i$ given $X_i$, and the outside operator is taken with respect to $X_i$. Thus, $\lbars(\thetavec)$ is consistent for $h(\thetavec)$, and $\lbar(\thetavec)$ and $\lbars(\thetavec)$ converge to the same function $h(\thetavec)$. Equivalently, the solutions of $\lbar'(\thetavec)=0$ and $\lbars'(\thetavec)=0$ converge to $\thetavec$, proving \textbf{Result~\ref{result:consistency}}.

\subsection{Result~\ref{res:normal}: Asymptotic Normality}
Consider the following Taylor series expansion,
$$0=\lbars'(\thetavechats)=\lbars'(\thetavechat)+\lbars''(\thetavechat)(\thetavechats-\thetavechat)+\cdots\,\,.$$
Thus,
\begin{align}\label{eqn:clt}
\sqrt{n}(\thetavechats-\thetavechat)|\Xvec_n\approx-[\lbars''(\thetavechat)]^{-1} [\sqrt{n}\,\lbars'(\thetavechat)].
\end{align}
Here $\lbars''(\thetavec)$ is the Hessian matrix. In the following, we will show that, conditional on $\Xvec_n$, $\lbars''(\thetavec)$ converges to the Fisher information matrix $I(\thetavec)$, and $\sqrt{n}\,\lbars'(\thetavechat)\to \N[0, \sigma^2I(\thetavec)]$, as $n\to\infty$. Note that $\sum_{i=1}^n l_i'(\thetavechat)=0$. Thus, $\sqrt{n}\,\lbars'(\thetavechat)$ can be written as
$$\sqrt{n}\,\lbars'(\thetavechat)=\frac{1}{\sqrt{n}}\sum_{i=1}^nZ_il_i(\thetavec; X_i)=\frac{1}{\sqrt{n}}\sum_{i=1}^n(Z_i-1)l_i'(\thetavechat).$$

In the following, we use $\Es$ and $\Vs$ to denote bootstrap expectation and variance. That is, these are moments determined by the iid exponential variables $Z_1,\ldots,Z_n$ treating the data $\Xvec_n$ as fixed. The bootstrap mean and variance of $\sqrt{n}\,\lbars'(\thetavechat)$ are obtained as
$$\Es\left[\sqrt{n}\,\lbars'(\thetavechat)\right]=\Es\left[\frac{1}{\sqrt{n}}\sum_{i=1}^n(Z_i-1)l_i'(\thetavechat)\right]=0,$$
and
\begin{align*}
\Vs\left[\sqrt{n}\,\lbars'(\thetavechat)\right]&=\Vs\left[\frac{1}{\sqrt{n}}\sum_{i=1}^n(Z_i-1)l_i'(\thetavechat)\right]=\frac{1}{n}\sum_{i=1}^n\left[l_i'(\thetavechat)\right]^2,
\end{align*}
respectively. Note that $\sum_{i=1}^n\left[l_i'(\thetavechat)\right]^2/n\to I(\thetavec)$. That is, the sample mean of squared score functions at the ML estimate converges in probability to the information matrix $I(\thetavec)$.

Hence, by using the weighted central limit theorem (i.e., the sum of independent, but weighted, exponential variables $Z_1,\ldots,Z_n$, given $\Xvec_n$),
\begin{align}\label{eqn:res1}
\sqrt{n}\,\lbars'(\thetavechat)\to \N\left[0, \sigma^2I(\thetavec)\right],
\end{align}
in which the convergence is in bootstrap probability given $\Xvec_n$. Also,
\begin{align}\label{eqn:res2}
\lbars''(\thetavechat)\to I(\thetavec).
\end{align}
in which the convergence is in bootstrap probability. That is because, conditional on $\Xvec_n$, $\lbars''(\thetavechat)$ has mean $\lbar{''}(\thetavechat)$ and variance $\sum_{i=1}^n [l_i{''}(\thetavec)]^2/n^2$ which goes to zero. Note that $\lbar''(\thetavechat)$ converges in probability to $I(\thetavec)$.

By \eqref{eqn:res1}, \eqref{eqn:res2}, and the Slutsky's theorem, we have $\sqrt{n}(\thetavechats-\thetavechat)|\Xvec_n\to \N\left[0,I(\thetavec)^{-1}\right]$, proving \textbf{Result~\ref{res:normal}}.

\subsection{Result~\ref{res:mle.exist}: Existence of ML Estimates}
Here we provide a proof of the result for the Weibull distribution. The proof of other commonly used log-location-scale distributions (i.e., lognormal, Fr\'{e}chet, and loglogistic distributions) is similar but lengthy, and thus it is omitted here. Here we prove the result under two particular cases. The pdf of the Weibull distribution is
\begin{align*}
f(t;\eta,\beta)=\frac{\beta}{\eta}\left(\frac{t}{\eta}\right)^{\beta-1}\exp\left[-\left(\frac{t}{\eta}\right)^\beta\right],\;t>0.
\end{align*}

For case (i) with two distinct failure times $t_1$ and $t_2$, we suppose that $t_1<t_2$, and $w_1$ and $w_2$ are the weights such that $w_1+w_2=1$. The weighted loglikelihood under the Weibull distribution is
\begin{align*}
l(\eta,\beta)&=\sum_{i=1}^{2}w_i\log \left[f\left(t_i;\eta,\beta\right)\right]=\sum_{i=1}^{2}w_i\left[\log (\beta)+(\beta-1)\log (t_i)-\beta\log(\eta)-\left(\frac{t_i}{\eta}\right)^\beta\right].
\end{align*}
Taking the derivative with respect to $\eta$ and setting it to 0, we obtain
\begin{align}\label{eqn:deri.eta}
\frac{\partial l}{\partial \eta}=\sum_{i=1}^{2}w_i\left(-\frac{\beta}{\eta}+\frac{\beta t_i^{\beta}}{\eta^{\beta+1}}\right)&=0.
\end{align}
Simplifying \eqref{eqn:deri.eta}, we obtain
\begin{align*}
-\frac{\beta}{\eta}+\frac{\sum_{i=1}^{2}w_i\beta t_i^\beta}{\eta^{\beta+1}}&=0.
\end{align*}
Solving the above equation, we obtain $\widehat{\eta}=\left(\sum_{i=1}^{2}w_it_i^\beta\right)^{\frac{1}{\beta}}$. Substituting $\widehat{\eta}$ into the loglikelihood, we obtain the log of the profile likelihood for $\beta$ as follows,
\begin{align*}
\widehat{l}(\beta)&=l(\widehat{\eta},\beta)=\sum_{i=1}^{2}w_i\left[\log (\beta)+(\beta-1)\log (t_i)-\log\left(\sum_{i=1}^{2}w_it_i^\beta\right)-\frac{t_i^\beta}{\sum_{i=1}^{2}w_it_i^\beta}\right].
\end{align*}
Taking the derivative with respect to $\beta$, we obtain,
$$
\frac{\partial \widehat{l}}{\partial \beta}=\frac{1}{\beta}+\sum_{i=1}^{2}w_i\log (t_i)-K,$$
where $K=\left[\sum_{i=1}^{2}w_i\log (t_i)t_i^\beta\right]/\left(\sum_{i=1}^{2}w_it_i^\beta\right).$ When $\beta\rightarrow 0$, $1/\beta\rightarrow\infty$, and the term $K$ is finite. Hence, $\partial\widehat{l}/\partial\beta\rightarrow \infty$. When $\beta\rightarrow\infty$, $1/\beta\rightarrow 0$, and the term
$$K\rightarrow\frac{w_2\log (t_2)t_2^\beta}{w_2t_2^\beta}=\log (t_2).$$
We obtain,
$$
\frac{\partial \widehat{l}}{\partial \beta}\rightarrow w_1\log (t_1)+w_2\log (t_2)-\log (t_2)=w_1[\log (t_1)-\log (t_2)],
$$
where $w_1[\log (t_1)-\log (t_2)]$ is a negative constant. Thus, $\partial\widehat{l}/\partial\beta=0$ has at least one  root.

To further show that $\widehat{l}$ has a global maximum, we restrict $\beta$ to a closed interval. Because $\lim_{\beta\rightarrow 0}\partial \widehat{l}/\partial \beta=\infty$, let $\tilde{\beta}_{\text{low}}=\displaystyle \min\{\beta|\beta\geq 0,\partial \widehat{l}/\partial \beta\leq C\}$ where $C$ is a large positive constant. Then the global maximum $\widehat{\beta}$ must be greater than or equal to $\tilde{\beta}_{\text{low}}$ because in the interval $(0,\tilde{\beta}_{\text{low}})$, $\widehat{l}$ is increasing. Because $\widehat{l}$ is continuous and $\lim_{\beta\rightarrow \infty}\partial\widehat{l}/\partial \beta=w_1[\log (t_1)-\log (t_2)]$, there exists $\tilde{\beta}_{\text{high}}$ such that $\forall\, \beta\geq\tilde{\beta}_{\text{high}}, \partial \widehat{l}/\partial \beta\leq 0.5w_1[\log (t_1)-\log (t_2)]$. Then $\widehat{\beta}$ must be less than or equal to $\tilde{\beta}_{\text{high}}$ because $\widehat{l}$ is decreasing when $\beta\geq\tilde{\beta}_{\text{high}}$. Then we have $\widehat{\beta}\in[\tilde{\beta}_\text{low},\tilde{\beta}_\text{high}]$. Because $\widehat{l}$ is continuous in a bounded interval $[\tilde{\beta}_\text{low},\tilde{\beta}_\text{high}]$, it must have a maximum in the closed interval, proving case (i).

For case (ii) with one failure time $t_1$ and a right-censored observation $t_2$ such that $t_2>t_1$, the weighted loglikelihood under the Weibull distribution is
\begin{align*}
l(\eta,\beta)&=w_1\log [f(t_1,\eta,\beta)]+w_2\log[1-F(t_2,\eta,\beta
)]\\
&=w_1\left[\log (\beta)+(\beta-1)\log (t_1)-\beta\log(\eta)-\left(\frac{t_1}{\eta}\right)^\beta\right]+w_2\left[-\left(\frac{t_2}{\eta}\right)^\beta\right].
\end{align*}
Taking the derivative with respect to $\eta$ and setting it to 0, we obtain,
\begin{align*}
\frac{\partial l}{\partial \eta}=
w_1\left(-\frac{\beta}{\eta}+\frac{\beta t_1^\beta}{\eta^{\beta+1}}\right)+
w_2\frac{\beta t_2^\beta}{\eta^{\beta+1}}=0.
\end{align*}
Solving the above equation, we obtain
$$
\widehat{\eta}=\frac{w_1t_1^\beta+w_2t_2^\beta}{w_1}.
$$
Substituting $\widehat{\eta}$ into $l(\eta,\beta)$, we obtain the log of the profile likelihood for $\beta$ as follows,
\begin{align*}
\widehat{l}(\beta)&=l(\widehat{\eta},\beta)\\&=w_1\left[\log (\beta)+(\beta-1)\log (t_1)-\log\left(\frac{\sum_{i=1}^{2}w_it_i^\beta}{w_1}\right)-\frac{w_1t_1^\beta}{\sum_{i=1}^{2}w_it_i^\beta}\right]
+w_2\left(\frac{w_1t_2^\beta}{\sum_{i=1}^{2}w_it_i^\beta}\right).
\end{align*}
Differentiating with respect to $\beta$ gives
\begin{align*}
\frac{\partial \widehat{l}}{\partial \beta}=
w_1\frac{1}{\beta}+w_1\log (t_1)-w_1K.
\end{align*}
Similar to case (i), one we can restrict $\widehat{\beta}$ to a closed interval $[\tilde{\beta}_{\text{low}}, \tilde{\beta}_{\text{high}}]$  and show that $\widehat{\beta}$ exists.

When the sample size $n$ is greater than 2, the proof is similar. The condition for the existence of the ML estimate will change slightly when $n>2$. For case (i), there should be at least two distinct failures. For case (ii), the condition is that there exists one right-censored observation which exceeds the failure time.


\end{document}